\documentclass[fleqn,usenatbib]{mnras}
\usepackage{caption}
\usepackage{graphicx}
\usepackage{caption} 
\usepackage{lipsum} 
\usepackage{newtxtext,newtxmath}
\usepackage{graphicx}
\usepackage{pdflscape}
\usepackage[normalem]{ulem}

\newcommand{\mmode}[1]{\ifmmode{#1}\else{$#1$}\fi}
\newcommand{\Teff}[0]{\mmode{T_\text{eff}}}
\newcommand{\Tsof}[0]{\mmode{T_\text{SOF}}}

\newcommand{\Mtot}[0]{\mmode{M_\text{tot}}}

\usepackage[T1]{fontenc}

\usepackage{graphicx}	
\usepackage{amsmath}	

\title[HdC Modelling]{Modelling Hydrogen-deficient Carbon stars in MESA--- The effects of total mass and mass ratio}

\author[C. Crawford et. al.]{Courtney L. Crawford$^{1}$\thanks{E-mail: courtney.crawford@sydney.edu.au},
Nikita Nikultsev$^{1}$,
Geoffrey C. Clayton$^{2,3}$,
Patrick Tisserand$^{4}$,
\newauthor Jamie Soon$^{5}$,
May G. Pedersen${^1}$
\\
$^{1}$Sydney Institute for Astronomy (SIfA), School of Physics, University of Sydney, NSW 2006, Australia\\
$^{2}$Department of Physics \& Astronomy, Louisiana State University, Baton Rouge, LA 70803, USA  \\
$^{3}$Space Science Institute (SSI), 4750 Walnut Street, Suite 205, Boulder, CO 80301, USA\\
$^{4}$Sorbonne Universit\'e, CNRS, UMR 7095, Institut d'Astrophysique de Paris, 98 bis bd Arago, 75014 Paris, France  \\ 
$^{5}$Research School of Astronomy and Astrophysics, Australian National University, Cotter Rd, Weston Creek ACT 2611, Australia \\
}

\date{Accepted XXX. Received YYY; in original form ZZZ}

\pubyear{2024}

\begin{document}
\label{firstpage}
\pagerange{\pageref{firstpage}--\pageref{lastpage}}
\maketitle

\begin{abstract}
Hydrogen-deficient Carbon (HdC) stars are rare, low-mass, chemically peculiar, supergiant variables believed to be formed by a double white dwarf (DWD) merger, specifically of a Carbon/Oxygen- (CO-) and a Helium-white dwarf (He-WD). They consist of two subclasses-- the dust-producing R Coronae Borealis (RCB) variables and their dustless counterparts the dustless HdCs (dLHdCs). Additionally, there is another, slightly cooler set of potentially related carbon stars, the DY Persei type variables which have some, but not conclusive, evidence of Hydrogen-deficiency. Recent works have begun to explore the relationship between these three classes of stars, theorizing that they share an evolutionary pathway (a DWD merger) but come from different binary populations, specifically different total masses (\Mtot{}) and mass ratios ($q$). In this work, we use the {\sc MESA} modelling framework that has previously been used to model RCB stars and vary the merger parameters, \Mtot{} and $q$, to explore how those parameters affect the abundances, temperatures, and luminosities of the resultant post-merger stars. We find that lower \Mtot{} and larger $q$'s both decrease the luminosity and temperatures of post-merger models to the region of the Hertzsprung-Russell Diagram populated by the dLHdCs. These lower \Mtot{} and larger $q$ models also have smaller oxygen isotopic ratios ($^{16}$O/$^{18}$O) which is consistent with recent observations of dLHdCs compared to RCBs. None of the models generated in this work can explain the existence of the DY Persei type variables, however this may arise from the assumed metallicity of the models. 
\end{abstract}





\section{Introduction}

Hydrogen-deficient Carbon (HdC) stars are an exotic class of star with Carbon and He-rich atmospheres, consisting of $\sim$1\% C, $\sim$98\% He and less than 1\% H by mass \citep{Clayton1996_HdCcomposition}. Despite having relatively low masses of $\sim$0.6 -- 1.2 \(M_\odot\) \citep{Clayton2012_HdCMasses}, HdC stars exhibit absolute V-band magnitudes of -2.6 to -5.2, comparable to classical supergiant stars of 12 \(M_\odot\) and above, and their temperatures are estimated to be $\sim$3500K--8000K \citep{Tisserand2022_dLHdC,Crawford2023_hdcclassification}. 
A very small number of HdC stars are much hotter with effective temperatures (\Teff{}) of about 20,000 K \citep{Clayton2012_HdCMasses}.
The HdC stars are believed to be formed via the merger of two white dwarfs \citep{Webbink1984_dwdmergerscenario}, and thus are quite old stars, found predominantly near old Galactic regions such as the bulge and the thick disk \citep{Tisserand2023_3Ddistgaia_ARXIVVERSION}. 
The HdC umbrella generally includes two sub-types of stars, the R Coronae Borealis (RCB) variables, and the dustless HdC (dLHdC) stars. 
The RCB stars are known for having aperiodic sudden drops in brightness, up to 8 magnitudes in the V-band, that can last from weeks up to years due the formation of dust at their surfaces \citep{Howell2013_dustmagnitudedrop8}. Conversely, the dLHdC stars, while having nearly identical optical spectra to the RCBs \citep{Crawford2023_hdcclassification}, do not show the same declines in brightness \citep{Tisserand2022_dLHdC}.
\citet{Tisserand2022_dLHdC} found that the dLHdC stars tend to be slightly warmer and less luminous than the RCB stars, which they theorized to be the result of slightly different merger populations, later supported by the population synthesis models included in \citet{Tisserand2023_3Ddistgaia_ARXIVVERSION}.
HdC stellar atmospheres exhibit rare isotopic ratios, including very large $^{12}$C/$^{13}$C and small $^{16}$O/$^{18}$O, the latter of which is taken as strong evidence for formation via double white dwarf merger \citep{Asplund2000_rcbabundances,Clayton2007_Excess018inRCBsandHDCsbecauseWDM,GarciaHernandez2010_oxygen18}. Recently, \citet{Karambelkar2022_oxygen18} found that observed dLHdC stars likely have more $^{18}$O in their atmospheres than RCB stars, which may support the case proposed in \citet{Tisserand2022_dLHdC}, that the dLHdC stars and RCB stars are formed via different merger populations.

Many works have also pointed out the strong similarities between the RCB variables and another class of variables, the DY Persei-type variables (DY Pers)  \citep{Alksnis1994_dyper,Keenan1997_DYPerseiTemperature,Zacs2007_dyper,Yakovina2009_dyper,Bhowmick2018_coolcousins,Crawford2023_hdcclassification,GarciaHernandez2023_dyper}. The similarities are both spectroscopic and photometric in origin. DY Pers are rare carbon stars with temperatures $<$3500K, where the prototype DY Persei, the warmest of its class, has an estimated \Teff{} of 2750-3500K \citep{Keenan1997_DYPerseiTemperature,GarciaHernandez2023_dyper}, and its optical spectrum looks similar to the coolest HdC star spectra \citep{Crawford2023_hdcclassification}. They are similar in brightness to the HdC stars, although slightly dimmer, at M$_V$~=~-1.5 to -3.0 mag \citep{Tisserand2009_MagellanicRCBsandDYPers}.
DY Pers show dust declines, however they tend to be shorter in length, shallower in depth, and more symmetrical than RCB declines \citep{Alcock2001_newmachorcb_dyper}. 
Evidence of H features (namely PAHs) have been observed in the DY Persei circumstellar dust cloud, similar to some extremely hot RCB stars, which may originate from a previous evolutionary phase or may point to a differing evolutionary pathway \citep{GarciaHernandez2013_dyperPAH}.
Historically, DY Per itself has been classified as a C-R:4 using the traditional carbon star classification system \citep{Barnbaum1996_carbonstaratlas,Keenan1997_DYPerseiTemperature}, indicating an observable CH-band, and thus presenting some H in its atmosphere. Due in part to the presence of this H, early works such as \citet{Tisserand2009_MagellanicRCBsandDYPers} suspected that DY Pers are more likely to be normal carbon stars that experience ejection events rather than an extension of the RCB phenomenon to lower temperature stars. However, \citet{Bhowmick2018_coolcousins} and \citet{GarciaHernandez2023_dyper} found that the DY Pers also have very small $^{16}$O/$^{18}$O ratios, just as the RCB and dLHdC stars do. \citet{GarciaHernandez2023_dyper} additionally explores DY Per's H-deficiency and while they cannot conclusively state that the star is H-deficient due to confusion with carbon bands or the F abundance contributing to the HF line, they show evidence of potential H-deficiency, which would make them another subclass of the HdC umbrella. The presence of $^{18}$O and the potential lack of H in the atmospheres of these stars is a likely indication that they may indeed share an evolutionary pathway with the RCB stars, i.e. through a double white dwarf merger.

RCB stars are believed to be formed by the merger of a Carbon/Oxygen (CO-) and a Helium white dwarf (He-WD) \citep{1984ApJS...54..335I}, and recent double WD (DWD) merger models have been able to reproduce many of the atmospheric conditions unique to RCB stars, including their temperatures and luminosities \citep{Weiss1987_reallyoldRCBevolution,Menon2013_ModifiedAGBevolutionOlder,Zhang2014_rcbmodels,Menon2019_AGBevolutionofRCBs,Schwab2019_2DintoMesaSimulations,Lauer2019_OriginalStellarEngineering,Crawford2020_RCBmodels,Munson2021_goingfrom3Dto1D,Munson2022_stellarengineeringpipeline,Gautschy2023_rcbpulsation_ARXIVVERSION,WongBildsten2024_rcbgyre}. In addition, recent works have posited that both the dLHdCs and the DY Per type variables are formed by different sets of initial DWD merger parameters, especially the total mass and mass ratio of these mergers (dLHdCs: \citealt{Tisserand2022_dLHdC,Karambelkar2022_oxygen18}, DY Pers: \citealt{Bhowmick2018_coolcousins,GarciaHernandez2023_dyper}).
In this work, we explore whether varying the initial parameters of an RCB model created using the DWD framework can instead be used to create models resembling either the dLHdC stars or the DY Per type variables, in both their temperatures and luminosities, as well as their slight differences in $^{18}$O abundances.
We focus on the main merger parameters-- the total post-merger mass (M$_{\rm tot}$) and the mass ratio ($q$~=~M$_{\rm donor}$/M$_{\rm accretor}$). Other possibly important parameters such as metallicity are also discussed. In Section~\ref{sec:pastwork} we review all previous work on DWD modelling. In Section~\ref{sec:modelling} we explain our modelling procedure, including any unique parameters we used. We present our results in Section~\ref{sec:results}, with Section~\ref{subsec:hrd} referring to the location of the models in \Teff{} and luminosity space, and Section~\ref{subsec:abunds} showcasing the effect on the surface abundances. Finally, we conclude and discuss in Section~\ref{sec:conclusion}, where we also provide our outlook on whether the dLHdC stars or the DY Pers can be explained by different initial merger parameters.


\section{Past Modelling Work}
\label{sec:pastwork}

Recent works have used a variety of different approaches to modelling HdC stars as post-DWD merger objects-- modified asymptotic giant branch (AGB) evolution \citep{Weiss1987_reallyoldRCBevolution, Menon2013_ModifiedAGBevolutionOlder, Menon2019_AGBevolutionofRCBs, Schwab2019_2DintoMesaSimulations, Gautschy2023_rcbpulsation_ARXIVVERSION}, accretion based models \citep{Saio2002_accretionrcbmodels,Zhang2014_rcbmodels}, hydrodynamical modelling \citep{Longland2011_hydrorcbmodel}, directly mapping a hydrodynamical merger model into 1D \citep{Schwab2019_2DintoMesaSimulations,Munson2021_goingfrom3Dto1D}, and most commonly, ``stellar engineered" models \citep{Lauer2019_OriginalStellarEngineering,Schwab2019_2DintoMesaSimulations, Crawford2020_RCBmodels,Munson2022_stellarengineeringpipeline,WongBildsten2024_rcbgyre}. A majority of these methods used the widely available stellar evolution code Modules for Experiments in Stellar Astrophysics \citep[MESA,][]{Paxton2011, Paxton2013, Paxton2015, Paxton2018, Paxton2019, Jermyn2023}. As we use the stellar engineering method in this work, we will solely discuss the literature surrounding this method.

The stellar engineering process is built from the method described in \citet{Shen2012_entropyinjection} and \citet{Shen2019_MESASS_dwdmerger}. It earned its name based on the fact that it builds a model resembling a post-merger object, rather than directly modelling the merger process through physical means such as accretion. The model generation process generally takes three steps. The first is to evolve a single star of mass equal to the desired \Mtot{} until it has a degenerate core. The evolution and physics are then stopped, and the model's abundances are changed to that of a core consistent with a CO-WD and an envelope consistent with a fully mixed He-WD, where the core/envelope boundary is determined by the mass ratio $q$. Finally, additional heating must be applied to the base of the envelope to account for the heat generated during the merger process, which creates the helium burning shell at the base of the envelope (called the ``Shell of Fire" or SOF in some works and T$_{\rm He}$ in others). Once the heat has been applied and the temperature-density profile resembles that of 3D hydrodynamical models, the model is then allowed to evolve normally within {\sc MESA}.

\citet{Lauer2019_OriginalStellarEngineering} was the first work to publish stellar engineered models for RCB stars. Their main set of models (the A Set) have \Mtot{}~=~0.8 M$_\odot$ and mass ratio $q$~=~0.45, and vary the initial radius of the post merger object, which is the same as varying the Helium burning shell temperature.
They also run five other models with higher total masses by changing the He-WD mass for a fixed CO-WD mass, thus varying both \Mtot{} and q.
Their work serves as a proof of concept, and finds good agreement with the location of the models within the Hertzsprung-Russel diagram (HRD), and general agreement with the surface abundances to observations, barring the high metallicity compared to RCB stars. Their models also show surface Li, which was later shown by \citet{Munson2021_goingfrom3Dto1D} to be an artifact of the choice of reaction network, as an inclusion of $^{11}$B in the reaction network destroys all surface lithium.

\citet{Schwab2019_2DintoMesaSimulations} created three types of models, originating from a 2D hydrodynamical model, from stellar engineering, and from modified He-star evolution. 
They use the AESOPUS web interface \citep{MarigoAringer2009_aesopus,Marigo2022_aesopus2} to create opacity tables which accommodate the abundance patterns from \citet{Weiss1987_reallyoldRCBevolution} and \citet{Asplund2000_rcbabundances}. These opacity tables are more appropriate for HdC stars than the defaults within {\sc MESA}. Additionally, they created a customized mass loss prescription for the RCB stars, rather than using the default {\sc MESA} wind prescription for AGB stars from \citet{Bloecker1995_massloss}.
The new opacities tend to move models towards the cool end of the HRD, and the mass loss predominantly effects the HdC phase lifetime and the mass of the star when it exits this phase.

\citet{Crawford2020_RCBmodels} created 18 models, 9 with solar metallicity and 9 with 10\% solar metallicity, using the stellar-engineering method, perturbing the temperature of the He-burning shell. These models all have \Mtot{}~=~0.8 $M_{\odot}$ and $q$~=~0.45. They found that higher He-burning shell temperatures (which they denote as $T_{\rm He}$) result in cooler RCB stars, and that the subsolar metallicity models produced warmer RCB stars. They also found that $T_{\rm He}$ plays a strong role in the surface abundances, and especially on the carbon and oxygen isotopic ratios. Additionally, they show that the surface abundances in RCB models are set very early on in their evolution, as the convective zone on the surface of the star disconnects from the processed material in the burning region many years before the models reach the HdC phase. They identify an ideal model with $T_{\rm He}$ of about 300 MK and subsolar metallicity. This model agrees with all observed RCB abundances except that of N.

While the work of \citet{Munson2021_goingfrom3Dto1D} was created by converting a 3D hydrodynamical model into a 1D RCB model, they also found very similar results to earlier stellar engineering models. Their models have \Mtot{}~=~0.8 $M_{\odot}$ and $q$~=~0.5, and they vary parameters such as the overshooting factor, which affects the boundaries of the convective regions, and initial H abundance in the envelope within {\sc MESA}. They also restricted themselves to either solar or 10\% solar metallicities. They found that some amount of overshooting was necessary to replicate the observed C/O ratio and the oxygen isotopic ratio (the no overshoot model has too much $^{18}$O). They also find little dependence of the surface abundances on the initial H abundance, although too much H results in disagreement with observations. Additionally, they performed nucleosynthesis in one extra model using the post-processing code NuGrid \citep{Herwig2008_nugrid,Pignatari2012_nugrid} rather than in {\sc MESA}. The intention for this NuGrid model was to test whether a larger reaction network could replicate the observed RCB enhancement of s-processed material, which their model fails to do due to not having enough H in the initial atmosphere.

\citet{Munson2022_stellarengineeringpipeline} focused on updating the stellar engineering process in two ways. The first is to change the functional form of the heat applied to the envelope during the model generation process to 
have a linear transition between the core and envelope regions, rather than an instantaneous discontinuity, 
which results in a smoother temperature-density profile for the post-merger object. The second is to utilize {\sc MESA}'s \texttt{one\_zone\_burn} functionality to simulate the nucleosynthesis that can happen during the DWD merger process. Their models all have 10\% solar metallicity, \Mtot{}~=~0.8 $M_{\odot}$, and $q$~=~0.5, though Model Set A experiments with the ``transition mass" (the outer mass coordinate of the the extra heat applied in the envelope). The new models created for this work now show enhancements in s-process material when using NuGrid, as opposed to the earlier models, and in fact are more consistent with the lesser known i-process \citep{Denissenkov2018_iprocess}.

Most recently, \citet{WongBildsten2024_rcbgyre} created stellar engineered RCB models with the intention of studying their pulsational properties using GYRE \citep{Townsend2013_gyre,Townsend2018_gyre,Goldstein2020_gyre}. They run three total models, one with \Mtot{}~=~0.7 $M_{\odot}$, $q$~=~0.4 and two with \Mtot{}~=~0.8 $M_{\odot}$, $q$~=~0.45. They do not include discussion on the surface abundances, however their models are well placed in the RCB region of the HRD. Their two lower \Mtot{} models are ran with the same initial conditions, only one with mass loss and one without, and show that the minimum temperature and luminosities in the HdC phase do not change, which is consistent with the results presented in \citet{Schwab2019_2DintoMesaSimulations} and \citet{Crawford2020_RCBmodels}.


\section{Merger Modelling}
\label{sec:modelling}

Our models are generated using the same method as \citet{Munson2022_stellarengineeringpipeline}, using {\sc MESA} version r21.12.1. For all models, we use the same CO- and He-WD abundances as \citet{Munson2022_stellarengineeringpipeline}, which are calculated using 10\% solar metallicity (scaled from GS98 \citealt{GS98_metalmixture}). We do not include rotation or overshooting. We use the reaction network created by \citet{Munson2021_goingfrom3Dto1D} called \texttt{rcb.net} which includes 40 relevant isotopes. We assume the thickness of the He-burning region to equivalent to 4\% of the CO core mass, as recommended by \citet{Munson2022_stellarengineeringpipeline}. Our models were all set to have the same He-burning region temperature (sometimes called the ``Shell of Fire", \Tsof{}) of 10$^{8.48 \pm 1}$ K, or 295-310 MK, which was found to be the ideal \Tsof{} in \citet{Crawford2020_RCBmodels} for replicating surface abundances.
We do not include mass loss in our models, following \citet{WongBildsten2024_rcbgyre}, as it has limited influence on the location of the models in \Teff{} and L space, and primarily influences the post-RCB evolution \citep{Schwab2019_2DintoMesaSimulations,Crawford2020_RCBmodels}. We do not use the customized HdC opacity tables from \citet{Schwab2019_2DintoMesaSimulations}, however we do use the ``Type 2" opacities from {\sc MESA}, which were created for high temperature, hydrogen-poor/metal-rich mixtures. Finally, we use a \texttt{mesh\_delta\_coeff} of 1 rather than 2 as was used in \citet{Munson2022_stellarengineeringpipeline}, which gives our models roughly twice as many spatial grid points, and aids our numerical convergence.

We run a grid of 42 total models varying only \Mtot{} and q, as listed in Table~\ref{tab:model_output}. A small portion of the grid was chosen to probe the effect of \Mtot{} around the region that most previous models focused, \Mtot{}~=~0.8 M$_\odot$ and q~=~0.45. Then we added lower \Mtot{} models while also varying the mass ratio. We focused our search towards the lower masses as the dLHdC stars and DY Pers tend to have lower luminosities than the RCBs, and in general stellar luminosity decreases with decreasing stellar mass.


\section{Results}
\label{sec:results}

We provide a summary of many output parameters for each of our models in Table~\ref{tab:model_output}. Our aim is to explore the effects of DWD merger parameters on post-merger supergiants, with the intent of disentangling the potential origins of the dLHdC stars or the DY Per stars. These two classes of stars are indeed very similar to the RCB stars, however there are minor differences that we hope to replicate with our models.

RCB stars are by far the most studied of these three classes of variable star. RCB stars have temperatures of approximately $\sim$3500K--8000K \citep{Tisserand2022_dLHdC,Crawford2023_hdcclassification}, and absolute V-band magnitudes of roughly -2.6 to -5.2 \citep{Tisserand2022_dLHdC}. We therefore define the ``RCB locus" of the HRD as log(L/L$_\odot$) from 3.6 to 4 and log(\Teff{}) from 3.6 to 3.9. Observed RCB stellar abundances come from \citet{Asplund2000_rcbabundances}. To summarize their abundances, RCB stars are known to have low metallicities of roughly -2 to -1 [Fe]\footnote{The term ``abundances" can have many meanings. Here we clarify our notation. We refer to abundances relative to solar abundances, which are defined for element `X' as [X]~=~log$\epsilon$(X) - log$\epsilon$(X)$_\odot$. log$\epsilon$(X) refers to the logarithm of the number of element `X' ions in a gas with 10$^{12.15}$ total ions. This is calculated as log$\epsilon$(X)~=~log$\frac{M_X}{\mu_X}$ + 12.15, where $M_X$ is the mass fraction of element X and $\mu_X$ is the mean atomic mass of element X.}$^{,}$\footnote{We note that RCB surface abundances are generally not reported as relative to H, as our atmospheres do not have strong H.}, along with enhanced CNO abundances, strongly enhanced F ([F] $\sim$ 2-3), and weakly enhanced s-process elements. They additionally have very high carbon isotopic ratios ($^{12}$C/$^{13}$C > 500) and weak oxygen isotopic ratios ($^{16}$O/$^{18}$O $\sim$ 1-10). Several RCBs show enhanced Li, which has not yet been explained, but has been discussed in \citet{Munson2021_goingfrom3Dto1D}. The RCB stars for which abundances have been measured in \citet{Asplund2000_rcbabundances} are biased towards the bright stars in the thick disk (see \citealt{Tisserand2023_3Ddistgaia_ARXIVVERSION} for Galactic kinematics of RCB stars) and thus their metallicities and abundances may not be true representations of the entire body of RCB stars.

dLHdC stars are found at HdC classes 4 and above \citep{Crawford2023_hdcclassification}, corresponding to roughly 4500-8500K in temperature. They are found ranging to lower luminosities than the RCB stars, M$_V$ from roughly -1.8 to -5.2 \citep{Tisserand2022_dLHdC}. We therefore define the ``dLHdC locus" of the HRD as log(L/L$_\odot$) from 3.2 to 4 and log(\Teff{}) from 3.75 to 3.9.
While not much work has been done regarding the dLHdC's abundances, as there were only five known until recently \citep{1967MNRAS.137..119W}, \citet{Kipper2002_dlhdcabunds_hmlib}'s abundance study of HD 137613 (HM Lib) has shown consistency with the range of RCB abundances found in \citet{Asplund2000_rcbabundances}. However, infrared studies of dLHdC stars have shown that they likely have more $^{18}$O than the RCB stars, and therefore a smaller oxygen isotopic ratio, $^{16}$O/$^{18}$O < 1 \citep{Karambelkar2022_oxygen18}. 
Finally, there is some evidence that dLHdCs may have weaker [N] and stronger [H] compared to RCBs, considering the overall weakness of their CN bands and the presence of H$\alpha$ lines in many dLHdC mid-resolution spectra \citep{Tisserand2022_dLHdC}.

DY Per type variables have temperatures less than 3500 K \citep{Keenan1997_DYPerseiTemperature,Tisserand2009_MagellanicRCBsandDYPers,GarciaHernandez2023_dyper,Crawford2023_hdcclassification}, and absolute V-band magnitudes of roughly -1.0 to -3.0 \citep{Tisserand2009_MagellanicRCBsandDYPers}. We therefore define the ``DY Per locus" as log(L/L$_\odot$) from 3.2 to 3.6 and log(\Teff{}) from 3.4 to 3.6. There are no comprehensive abundance studies of the DY Per type variables, however \citet{Bhowmick2018_coolcousins} and \citet{GarciaHernandez2023_dyper} have studied the $^{18}$O using infrared spectra and found that the DY Per type variables have oxygen isotopic ratios consistent with RCB stars, roughly $^{16}$O/$^{18}$O $>$ 5. \citet{GarciaHernandez2023_dyper}'s study of DY Per itself showed that the star is not H-rich, but cannot conclusively state whether the star is truly H-deficient to the level of RCB stars. \citet{Zacs2007_dyper}'s study of DY Per's optical band spectra estimated the star to have roughly solar [Fe] and [s/Fe], contrary to typical RCB stars which have [Fe] between -2 to -1 and enhanced s-processed material on their surfaces.

\onecolumn
\begin{landscape}
\begin{table*}
    \caption{All model observable parameters}
    \label{tab:model_output}
    \begin{tabular}{ccccccccccccccc}
        M$_{\text{total}}$ & $q$ & log(T$_{\text{eff,HdC}}$) & log(L$_{\rm HdC}$) & log(R$_{\rm HdC}$) & HdC Lifetime & \% lifetime & log$\epsilon$(C) & log$\epsilon$(N) & log$\epsilon$(N) & log$\epsilon$(O) & log$\epsilon$(F) & C/O & $^{16}\text{O}/^{18}\text{O}$ & $^{12}\text{C}/^{13}\text{C}$\\
        (M$_{\odot}$) & & (K) & (L$_{\odot}$) & (R$_{\odot}$) & (10$^5$ yr) & in HdC phase & & (at HdC start) & (at HdC end) & & & & & ($\times$ 10$^6$)\\
        \hline
0.675 & 0.525 & 3.77 & 3.69 & 1.82 & 1.20 & 21.32 & 9.21 & 7.36 & 7.96 & 8.60 & 6.07 & 3.06 & 19.64 & 7.30 \\
0.675 & 0.550 & 3.77 & 3.65 & 1.81 & 1.04 & 16.88 & 9.27 & 7.39 & 7.95 & 8.60 & 6.30 & 3.54 & 27.54 & 7.44 \\
0.675 & 0.625 & 3.78 & 3.55 & 1.73 & 0.70 & 9.20 & 9.27 & 7.48 & 7.95 & 8.60 & 6.25 & 3.45 & 20.08 & 7.55 \\
0.675 & 0.650 & 3.79 & 3.52 & 1.70 & 0.64 & 7.91 & 9.24 & 7.54 & 7.95 & 8.61 & 6.15 & 3.24 & 16.99 & 7.33 \\
0.675 & 0.675 & 3.80 & 3.49 & 1.67 & 0.55 & 6.41 & 9.22 & 7.52 & 7.95 & 8.61 & 6.04 & 3.07 & 13.99 & 7.45 \\
0.700 & 0.400 & 3.74 & 3.92 & 2.01 & 2.14 & 64.38 & 9.25 & 7.06 & 7.96 & 8.59 & 6.21 & 3.47 & 46.43 & 7.91 \\
0.700 & 0.450 & 3.75 & 3.86 & 1.95 & 2.41 & 62.36 & 9.22 & 7.16 & 7.95 & 8.60 & 6.29 & 3.18 & 36.11 & 7.17 \\
0.700 & 0.500 & 3.76 & 3.78 & 1.90 & 2.63 & 55.58 & 9.24 & 7.26 & 7.95 & 8.60 & 6.31 & 3.31 & 29.39 & 7.75 \\
0.700 & 0.550 & 3.77 & 3.72 & 1.85 & 0.71 & 12.54 & 9.24 & 7.32 & 7.95 & 8.60 & 6.19 & 3.28 & 24.76 & 7.83 \\
0.700 & 0.575 & 3.77 & 3.69 & 1.82 & 1.75 & 30.27 & 9.21 & 7.35 & 7.95 & 8.60 & 6.16 & 3.05 & 21.32 & 7.34 \\
0.725 & 0.350 & 3.71 & 4.05 & 2.13 & 1.84 & 75.33 & 9.32 & 6.94 & 7.94 & 8.59 & 6.57 & 4.03 & 158.80 & 10.14 \\
0.725 & 0.400 & 3.73 & 3.99 & 2.06 & 2.24 & 75.30 & 9.24 & 7.02 & 7.96 & 8.59 & 6.12 & 3.35 & 51.73 & 8.01 \\
0.725 & 0.450 & 3.75 & 3.92 & 1.98 & 2.59 & 75.12 & 9.16 & 7.10 & 7.97 & 8.59 & 6.00 & 2.78 & 24.87 & 7.32 \\
0.725 & 0.500 & 3.75 & 3.85 & 1.95 & 3.15 & 74.55 & 9.24 & 7.23 & 7.95 & 8.59 & 6.41 & 3.34 & 39.29 & 8.19 \\
0.725 & 0.550 & 3.77 & 3.78 & 1.88 & 3.52 & 72.76 & 9.19 & 7.27 & 7.96 & 8.60 & 6.08 & 2.89 & 20.36 & 7.88 \\
0.750 & 0.350 & 3.70 & 4.10 & 2.17 & 1.82 & 81.60 & 9.33 & 6.90 & 7.95 & 8.59 & 6.55 & 4.12 & 198.44 & 11.53 \\
0.750 & 0.400 & 3.73 & 4.04 & 2.09 & 2.21 & 82.27 & 9.24 & 6.97 & 7.97 & 8.59 & 6.11 & 3.40 & 57.23 & 8.85 \\
0.750 & 0.450 & 3.75 & 3.97 & 2.01 & 2.57 & 82.12 & 9.16 & 7.03 & 7.97 & 8.59 & 5.94 & 2.78 & 28.55 & 7.50 \\
0.750 & 0.500 & 3.74 & 3.91 & 2.00 & 3.17 & 83.19 & 9.24 & 7.15 & 7.95 & 8.59 & 6.50 & 3.37 & 49.04 & 8.92 \\
0.750 & 0.550 & 3.76 & 3.84 & 1.92 & 3.61 & 83.27 & 9.18 & 7.24 & 7.96 & 8.60 & 6.13 & 2.91 & 25.07 & 8.08 \\
0.775 & 0.350 & 3.70 & 4.16 & 2.21 & 1.77 & 85.88 & 9.31 & 6.94 & 7.95 & 8.58 & 6.76 & 3.98 & 157.41 & 12.68 \\
0.775 & 0.400 & 3.72 & 4.09 & 2.13 & 2.13 & 86.40 & 9.26 & 6.96 & 7.96 & 8.58 & 6.56 & 3.58 & 90.84 & 10.62 \\
0.775 & 0.425 & 3.72 & 4.07 & 2.13 & 2.36 & 86.60 & 9.26 & 7.07 & 7.96 & 8.59 & 6.37 & 3.54 & 82.13 & 9.54 \\
0.775 & 0.550 & 3.76 & 3.90 & 1.96 & 3.48 & 88.34 & 9.19 & 7.16 & 7.97 & 8.59 & 6.09 & 2.93 & 30.36 & 8.24 \\
0.775 & 0.575 & 3.76 & 3.87 & 1.93 & 3.69 & 88.59 & 9.16 & 7.19 & 7.97 & 8.60 & 6.01 & 2.71 & 23.17 & 7.89 \\
0.700 & 0.625 & 3.78 & 3.62 & 1.77 & 1.15 & 17.34 & 9.22 & 7.45 & 7.95 & 8.60 & 6.11 & 3.08 & 18.50 & 7.50 \\
0.700 & 0.650 & 3.79 & 3.59 & 1.74 & 0.95 & 13.40 & 9.21 & 7.49 & 7.95 & 8.61 & 6.13 & 3.04 & 17.56 & 7.41 \\
0.725 & 0.625 & 3.78 & 3.68 & 1.81 & 3.61 & 61.21 & 9.21 & 7.39 & 7.95 & 8.60 & 6.18 & 3.03 & 21.70 & 7.42 \\
0.725 & 0.650 & 3.79 & 3.66 & 1.78 & 2.15 & 34.45 & 9.18 & 7.42 & 7.96 & 8.61 & 6.08 & 2.84 & 17.40 & 7.41 \\
0.750 & 0.625 & 3.78 & 3.75 & 1.85 & 4.38 & 82.85 & 9.18 & 7.35 & 7.96 & 8.60 & 6.16 & 2.82 & 21.05 & 7.60 \\
0.750 & 0.650 & 3.78 & 3.72 & 1.83 & 4.63 & 82.48 & 9.18 & 7.38 & 7.96 & 8.60 & 6.11 & 2.83 & 20.52 & 7.49 \\
0.775 & 0.625 & 3.77 & 3.80 & 1.89 & 4.27 & 88.88 & 9.18 & 7.29 & 7.96 & 8.60 & 6.14 & 2.82 & 24.25 & 7.81 \\
0.775 & 0.650 & 3.77 & 3.78 & 1.86 & 4.53 & 89.08 & 9.17 & 7.34 & 7.95 & 8.60 & 6.35 & 2.80 & 23.65 & 7.90 \\
0.725 & 0.450 & 3.74 & 3.92 & 2.00 & 2.60 & 75.38 & 9.22 & 7.09 & 7.96 & 8.59 & 6.27 & 3.14 & 44.47 & 7.28 \\
0.738 & 0.450 & 3.75 & 3.95 & 2.01 & 2.60 & 78.94 & 9.19 & 7.06 & 7.96 & 8.60 & 6.16 & 2.96 & 39.54 & 7.09 \\
0.760 & 0.450 & 3.74 & 4.00 & 2.04 & 2.54 & 84.42 & 9.19 & 7.04 & 7.96 & 8.59 & 6.20 & 3.00 & 43.39 & 7.82 \\
0.770 & 0.450 & 3.74 & 4.02 & 2.06 & 2.50 & 86.37 & 9.19 & 7.01 & 7.97 & 8.59 & 6.32 & 2.98 & 48.00 & 8.11 \\
0.790 & 0.450 & 3.73 & 4.06 & 2.08 & 2.41 & 88.98 & 9.19 & 7.00 & 7.97 & 8.59 & 6.32 & 2.99 & 39.65 & 8.55 \\
0.805 & 0.450 & 3.73 & 4.08 & 2.11 & 2.33 & 90.61 & 9.22 & 6.98 & 7.97 & 8.59 & 6.44 & 3.16 & 58.53 & 9.05 \\
0.818 & 0.450 & 3.72 & 4.11 & 2.13 & 2.27 & 91.28 & 9.22 & 6.96 & 7.96 & 8.59 & 6.59 & 3.22 & 62.45 & 9.86 \\
0.830 & 0.450 & 3.72 & 4.13 & 2.15 & 2.21 & 92.77 & 9.22 & 6.99 & 7.95 & 8.59 & 6.74 & 3.21 & 67.08 & 10.68 \\
0.850 & 0.450 & 3.72 & 4.17 & 2.18 & 2.12 & 93.40 & 9.23 & 7.02 & 7.93 & 8.59 & 6.94 & 3.29 & 60.40 & 12.90 \\
\hline
\multicolumn{5}{c}{RCB majority \citep{Asplund2000_rcbabundances}}  &  &  & 7.7-8.9 & \multicolumn{2}{c}{8.3-9.1}& 7.5-9.0 & 6.9-7.2 & $\gtrapprox$ 1 & $\sim$1-10 & $>$500 \\
\multicolumn{5}{c}{Solar \citep{Lodders2003_SolarAbundances}}  &  &  & 8.46 & \multicolumn{2}{c}{7.90} & 8.76 & 4.53 & 0.5 & 500 & 89\\
\hline
\multicolumn{15}{l}{Note. We quote the surface abundances as not relative to solar here to remove the dependence on assumed solar abundances for easier future comparisons.} \\
    \end{tabular}
\end{table*}
\end{landscape}
\twocolumn


\subsection{HRD Location}
\label{subsec:hrd}

For all of our post-merger models, we define the phase of interest, the ``HdC phase", as the position of their minimum \Teff{}, which roughly corresponds to the peak luminosity.
For the purposes of lifetime calculation, we define the end of the HdC phase as the time when a model's log(\Teff{}) rises to >~3.9.
The majority of models spend the longest amount of evolution time 
at their minimum \Teff{},
though this is a function of mass. The percentage of each model's lifetime spent in the HdC phase is listed in Table~\ref{tab:model_output}. This lifetime is also a strong function of the assumed mass loss \citep{Crawford2020_RCBmodels}, which is not well understood. Generally, models spend more than 85\% of the giant phase lifetime in the HdC phase. Therefore, if we were to observe these as real stars, this is the phase in which we'd be most likely to observe them.
Their surface abundances additionally do not change through this phase, as their surface convection zones have disconnected from their He-burning shells \citep{Crawford2020_RCBmodels}. The one exception is the abundances of $^{14}$C and $^{14}$N as the former radioactively decays, making these stars potentially interesting subjects for carbon dating \citep{Munson2022_stellarengineeringpipeline}. The internal structure does change past this point, however this could only be studied using asteroseismology \citep{WongBildsten2024_rcbgyre}. We therefore focus all further analysis solely on each model's HdC phase location and abundances.

\begin{figure}
    \centering
    \includegraphics[width=\columnwidth]{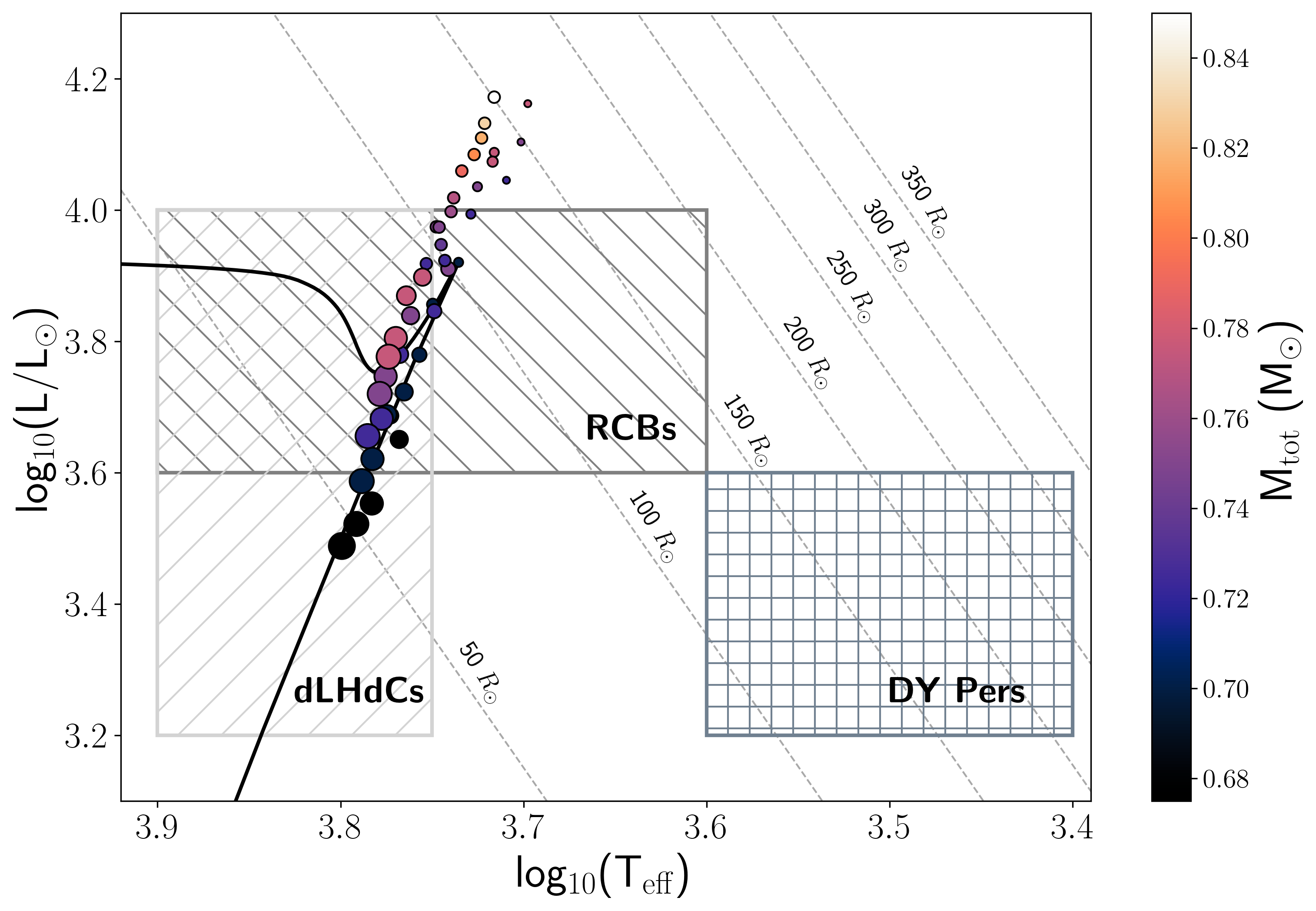}
    \caption{HR diagram for all models. In the hatched regions, we show the loci for the RCBs, the dLHdCs, and the DY Pers. Note that the loci for the RCBs and the dLHdCs overlap. We plot the location of the HdC phase for each of our 42 models in circles, where color represents the \Mtot{} and the size of the point represents the mass ratio $q$ (smaller point denotes a smaller $q$). The black line shows a typical evolutionary track for one of our models (\Mtot{}~=~0.7, $q$~=~0.4). We also show diagonal lines of constant radius.
    }
    \label{fig:hrd_all}
\end{figure}

\begin{figure}
    \centering
    \includegraphics[width=\columnwidth]{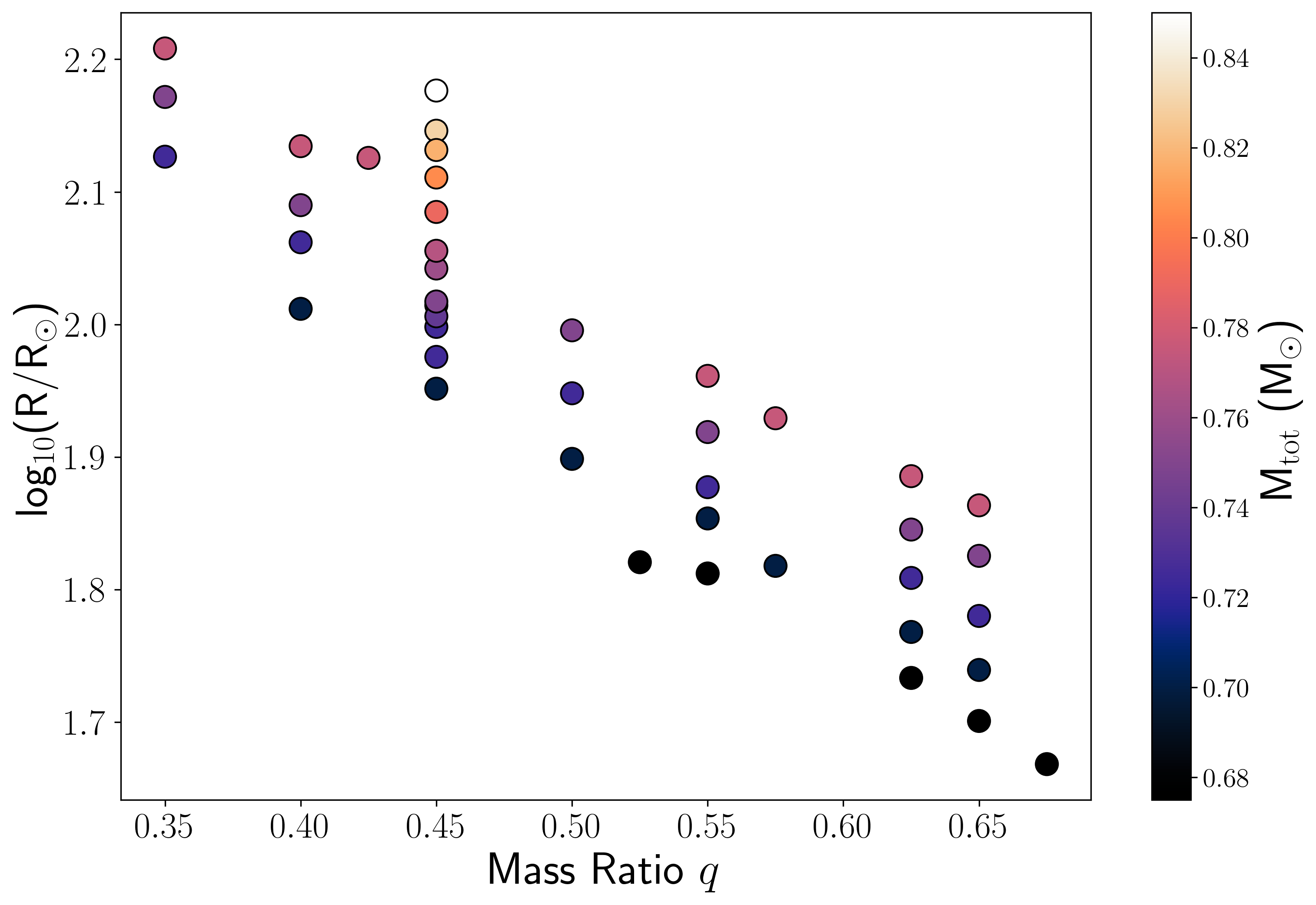}
    \caption{The radius versus the mass ratio $q$ for each of our models, colored by their \Mtot{}.}
    \label{fig:radius_v_q}
\end{figure}

\begin{figure}
    \centering
    \includegraphics[width=\columnwidth]{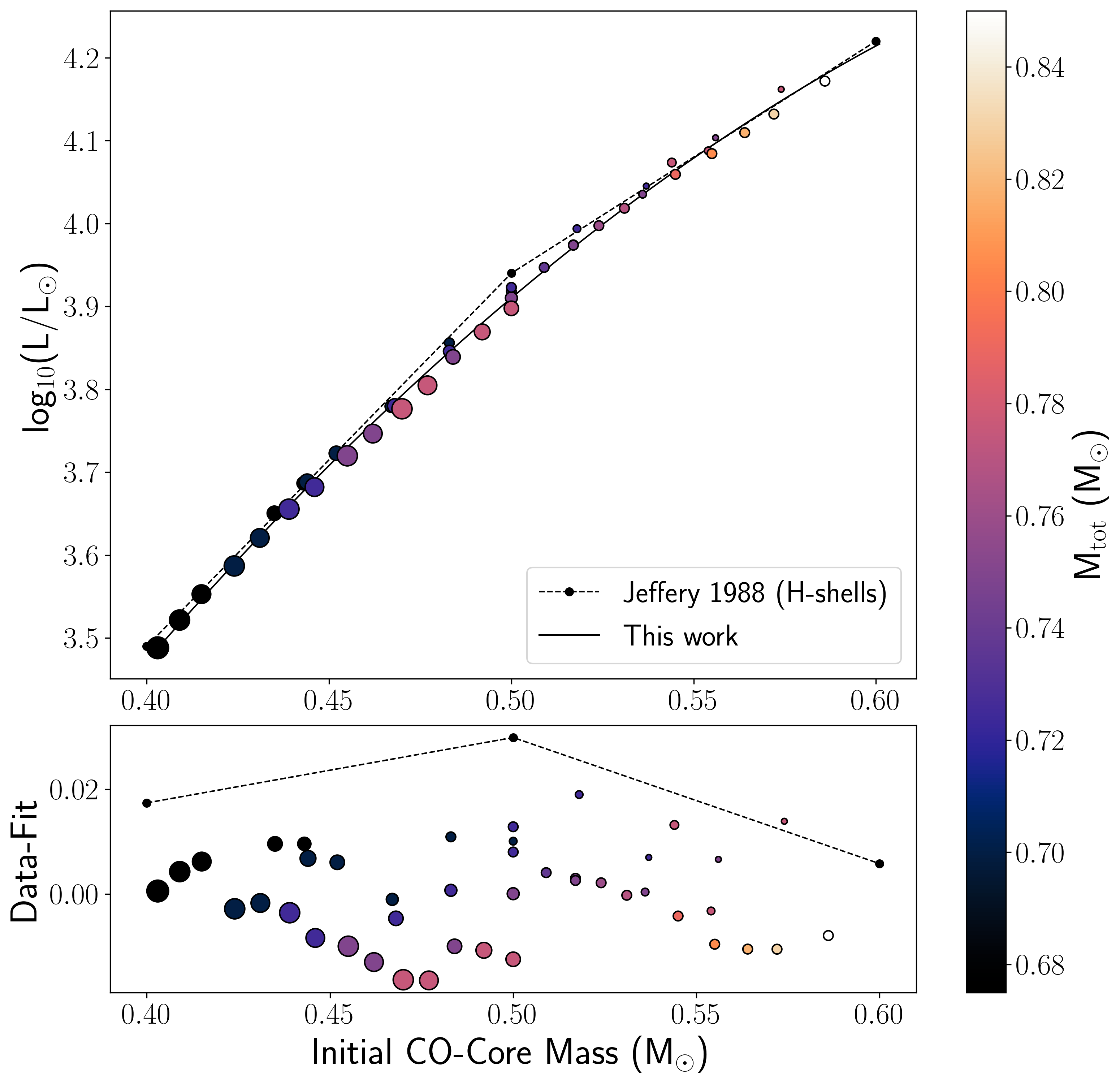}
    \caption{The core mass-luminosity relationship for all models. All models are colored according to their \Mtot{} and the size of the marker increases with increasing mass ratio $q$. We show the relation found for H-burning shells in \protect\citet{Jeffery1988_coremassluminosity} in the dashed line. In the solid black line, we show our quadratic fit to this relation, with the functional form $L_{\rm HdC} = -6.67M_{\rm core}^2 + 0.48M_{\rm core} + 3.9$. In the lower panel we show the data minus the quadratic fit, for both the model points and the relation from \protect\citet{Jeffery1988_coremassluminosity}.}
    \label{fig:rcbcoremass}
\end{figure}

We show the HRD position of each model in Figure~\ref{fig:hrd_all}, colored by \Mtot{} and with size increasing with increasing mass ratio. In different hatched regions we mark the RCB, dLHdC, and DY Per loci. In general, we see that models with lower \Mtot{} and higher $q$ tend to have lower luminosities and higher temperatures, which is consistent with a change in radius as shown in Figure~\ref{fig:radius_v_q}, and is inversely proportional to the change in surface \Teff{}. In fact the lowest \Mtot{} and highest $q$ models in this grid are consistent with the location of the dLHdC locus. 
A similar relationship between luminosity and merger parameters is also seen in other RCB merger models, such as \citet{SaioJeffery2000_mergermodels} and \citet{Zhang2014_rcbmodels}. The true nature of this relationship is a connection between the star's core mass and the luminosity of the burning shell encircling it \citep{Paczynski1970_coremassluminosity,Jeffery1988_coremassluminosity,Saio1988_coremassluminosity,WongBildsten2024_rcbgyre}. We show the CO-core mass - luminosity relation in Figure~\ref{fig:rcbcoremass}, to which we fit a quadratic function, such that $L_{\rm HdC} = -6.67M_{\rm core}^2 + 0.48M_{\rm core} + 3.9$. Note that the CO-core mass of each model at the onset of RCB phase has not significantly changed from its initial value as it has not yet had sufficient time to grow.
The higher \Mtot{} models are the most similar to previous stellar engineered models, which usually have \Mtot{}~=~0.8 M$_\odot$ and $q$~=~0.45. All of these models have high luminosities, log(L/L$_\odot$) > 3.8. This is consistent with all other works which employ stellar engineered models, as no published work using this kind of model shows an HdC phase with log(L/L$_\odot$) < 3.8 \citep{Lauer2019_OriginalStellarEngineering,Schwab2019_2DintoMesaSimulations,Crawford2020_RCBmodels,Munson2021_goingfrom3Dto1D,Munson2022_stellarengineeringpipeline,WongBildsten2024_rcbgyre}. These models span a narrow range of log(\Teff{}) from $\sim$ 3.7 -- 3.75. Although our helium burning shell temperature is consistent, our HdC phase \Teff{} is slightly cooler than the corresponding model from \citet[][ Model SUB8.48, log(\Teff{})=3.88]{Crawford2020_RCBmodels}, due to the slight differences in the model generation routine made in \citet{Munson2022_stellarengineeringpipeline}. In general, all the models span a limited range of log(\Teff{}) from 3.7 to 3.8, or \Teff{} from 5000 to 6300 K. Previous works have shown that the HdC phase \Teff{} of these models can be changed by changing the \Tsof{}, metallicity, or opacity tables \citep{Schwab2019_2DintoMesaSimulations,Crawford2020_RCBmodels}, and therefore a diverse range of DWD mergers could fill in the full \Teff{} range of the RCB locus.

\subsection{Abundances}
\label{subsec:abunds}

As discussed at the beginning of Section~\ref{sec:results}, only the RCB star abundances have been studied, although those analyses are biased towards bright RCBs. There have not been any focused abundance studies of the dLHdCs (although a study of one star shows consistency with RCBs, \citealt{Kipper2002_dlhdcabunds_hmlib}) or the DY Pers. Therefore, due to their optical spectroscopic similarities we assume the latter two classes have similar overall abundances to the RCB stars, except for the known differences in oxygen isotopic ratios. We therefore compare all of our models to the abundances measured in \citet{Asplund2000_rcbabundances}.

Across all of our models, we see very little change in the surface abundances of Li, C, O, Ne, Na, and Mg, likely due to us fixing the \Tsof{} to a single value. See \citet{Crawford2020_RCBmodels} for discussion of each of these major elements in stellar engineered models. We list the main surface abundances of all models in Table~\ref{tab:model_output} along with the observed abundance ranges from \citet{Asplund2000_rcbabundances} and the solar values measured by \citet{Lodders2003_SolarAbundances}.
Elements O, Ne, Na, and Mg are consistent with observed RCB abundances. Our models have enhanced C such that all models have C/O ratios > 1, and therefore may all be considered carbon stars. 
The C on the surface of our models is higher than reported observations on stars, however there is a known issue with HdC stars dubbed the ``carbon problem" \citet{Asplund2000_rcbabundances} where the C surface abundances measured are not self-consistent, and must be instead assumed. Therefore, the disagreement of our model surface C with observations may not be critical.
Our models do not show any surface lithium, and the existence of surface Li in RCB stars is a known problem, which cannot be easily explained with any stellar engineered models, even those with post-processed abundances (see \citealt{Munson2021_goingfrom3Dto1D} for discussion). In literature, only four RCB stars have measured Li abundances (R CrB, UW Cen, RZ Nor, and SU Tau), however the recent discovery of many new RCB stars and dLHdC stars has facilitated potential detection of Li lines in low resolution spectra for 17 dLHdC stars and 18 RCB stars \citep{Crawford2023_hdcclassification}. 
All models show strong carbon isotopic ratios. The minimum model $^{12}$C/$^{13}$C is 7.09~$\times$~10$^6$, which is consistent with the observed lower bound of $\sim$~500.

\begin{figure}
    \centering
    \includegraphics[width=\columnwidth]{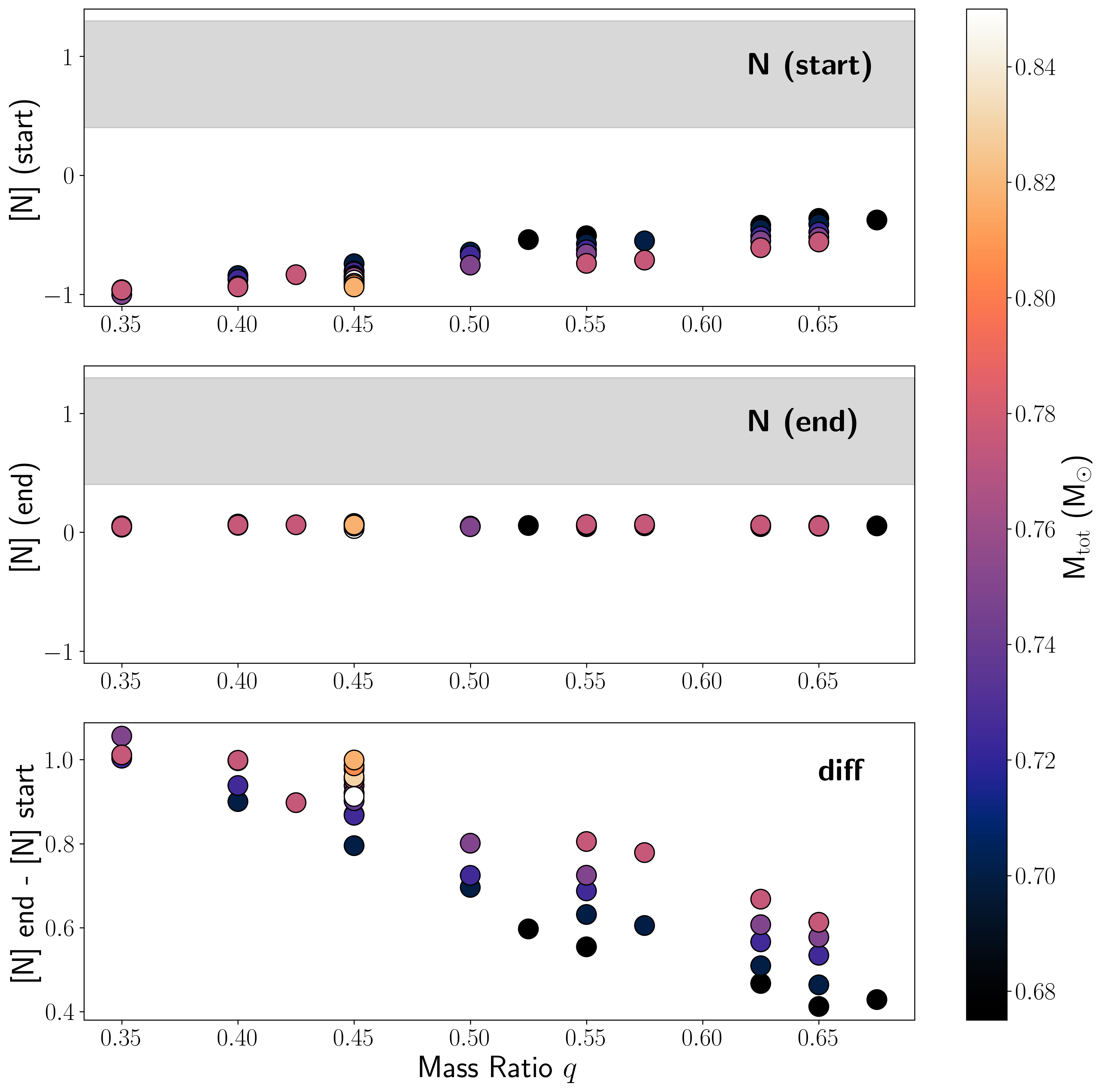}
    \caption{Here we plot three representations of the N abundances scaled to solar values ([N]) for our models. The top panel shows [N] at the beginning of HdC phase, the middle panel shows [N] at the end of the HdC phase, and the lowest panel shows the difference between the two times. The upper two panels have a grey shaded region which denotes the observed [N] abundance range for known RCB stars. Each is plotted against the mass ratio $q$ and colored by the \Mtot{}.
    }
    \label{fig:abund_N}
\end{figure}

[N] is also a known problem for stellar engineered models which have not been able to replicate the elevated abundances observed in RCBs \citep{Crawford2020_RCBmodels,Munson2021_goingfrom3Dto1D,Munson2022_stellarengineeringpipeline}. However, \citet{Crawford2020_RCBmodels} shows that the surface [N] is a strong probe of the triple-alpha reaction chain $^{14}N(\alpha, \gamma) ^{18}F(\beta^+) ^{18}O(\alpha, \gamma) ^{22}Ne$, and \citet{Munson2022_stellarengineeringpipeline} have shown that [N] is strongly affected by the neutron poison reaction $^{14}$N(n,p)$^{14}$C and subsequent decay from $^{14}$C back into $^{14}$N. It is possible that the elevated [N] we observe in RCB stars is due to this latter process, where the $^{14}$C decays with a half-life of 5730 $\pm$ 40 yr. As the HdC phase lifetimes of these models are on the order of 10$^5$ years (see Table~\ref{tab:model_output}), we should expect to see the effect of this decay on the [N] abundances over time (though not on [C], as $^{14}$C is not a significant isotope of C compared to $^{12}$C).
In fact, this decay is enough to bring the stellar engineered models into agreement by the end of the HdC phase lifetime in the work of \citet{Munson2022_stellarengineeringpipeline}. 
In Figure~\ref{fig:abund_N} we show the [N] surface abundances at the start and end of HdC phase for our models in comparison with observations, as well as the difference between the start and end values.
In our dataset we see that our models have initial [N] from -1.0 to -0.2 dex, increasing with increasing q. Models with the same $q$ have roughly the same [N] abundance, with a small ($\sim$0.1 dex) increase for lower \Mtot{} models. However, the final [N] as the star leaves the HdC phase is roughly constant across all models at [N]~=~0.05 dex, which is much closer to the range of RCB observations, but not quite in agreement.
The value of surface [N] at the end of the HdC phase is diminished from the model's initialization value by $\sim$ 0.4 dex.
In the lowest panel we can see that the lowest $q$ models have the largest change in [N] over the course of the HdC phase evolution at roughly 1.0 dex, which decreases with q, and also decreases for lower \Mtot{} models at the same q. The final [N] abundance is likely constant due to all models having the same initial He-burning region temperatures.
Given that [N] is constant across all models, it is difficult to interpret why the dLHdCs should have weaker CN bands, as found by \citet{Tisserand2022_dLHdC}. The apparent lower [N] abundance in dLHdCs could perhaps be explained by a change in the He-burning shell temperature (T$_{\rm SOF}$) from low-mass mergers, or from differing initial [N] abundances from the progenitor He-WD.

\begin{figure}
    \centering
    \includegraphics[width=\columnwidth]{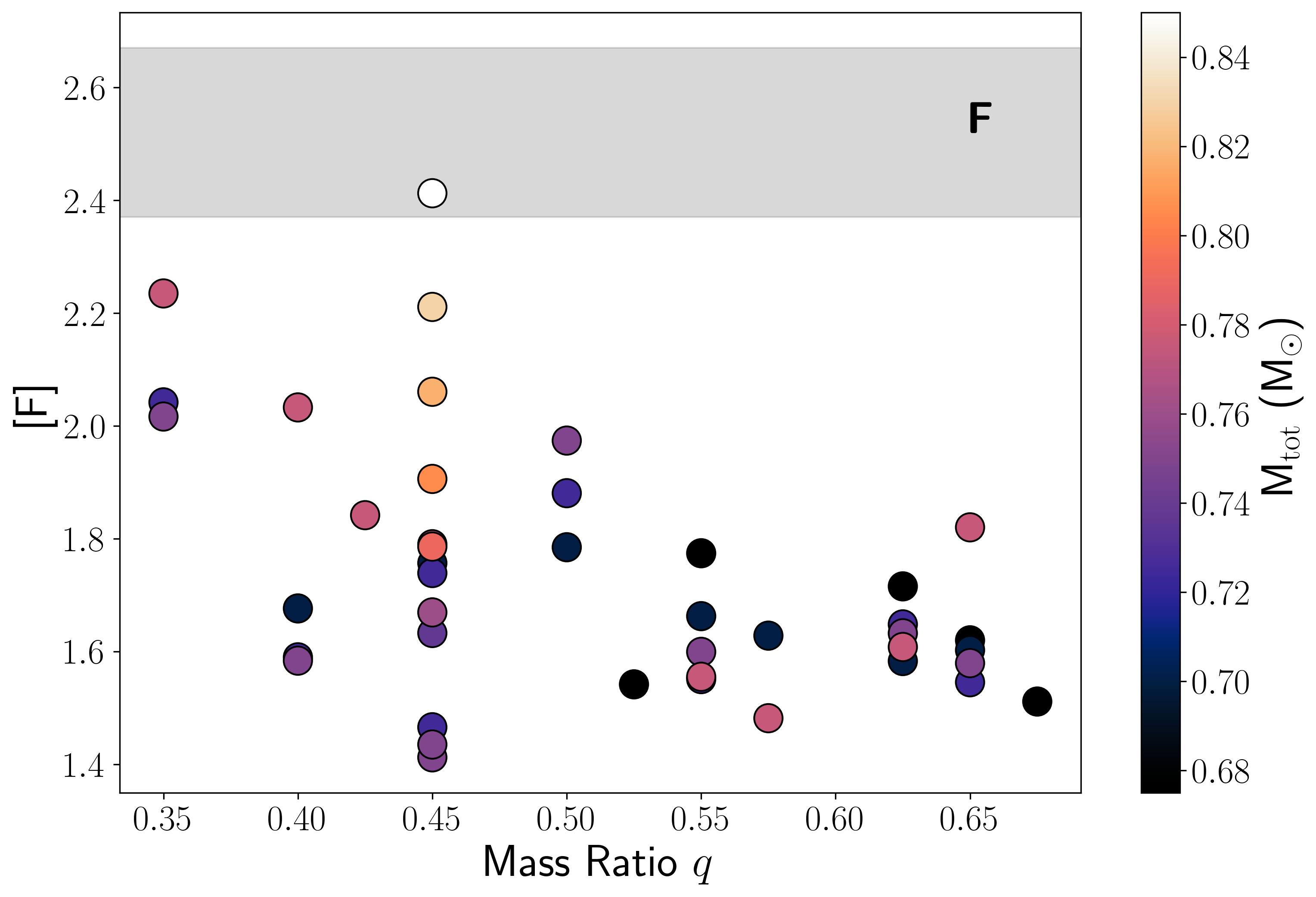}
    \caption{Here we show the F abundance for each model scaled to the solar value (denoted [F]), plotted versus the mass ratio $q$ and colored by the \Mtot{}. We also show a grey shaded region which denotes the observed [F] abundance range for known RCB stars.
    }
    \label{fig:abund_F}
\end{figure}

We also show the [F] surface abundances for our models in Figure~\ref{fig:abund_F}.
F, and in particular the stable isotope $^{19}$F, is a strong tracer element for the isotope $^{18}$O, as the reaction chain $^{14}N(\alpha, \gamma) ^{18}F(\beta^+)^{18}O$ can result in a p-capture $^{18}$O(p,$\gamma$)$^{19}$F, though $^{19}$F can also be formed via direct $\alpha$-capture on $^{15}$N.
The abundance of [F] can only be measured for particularly hot RCB stars, but is known to be strongly enhanced in the stars for which it has been measured \citep{Asplund2000_rcbabundances,Bhowmick2020_FluorineinEHeStars}. Additionally, from the analysis of \citet{GarciaHernandez2023_dyper}, the strength of the HF band in DY Per may be consistent with H-deficiency and F-enhancement, as is seen in the RCB stars. However, due to the lack of [F] abundance measurements in real stars, the range of observed abundances is very small. All models in our sample show strongly enhanced [F] of nearly +2 dex, which is especially strong considering the input [Fe] of each model is -1 dex. In general, we see a decrease in surface [F] with an increase in q, and some inconsistent variation in different \Mtot{} values. While only the highest \Mtot{} model appears consistent with observations, we emphasize that the observational data for this is largely incomplete, and is based on only four stars which are warm enough for surface [F] to be measured.

\begin{figure}
    \centering
    \includegraphics[width=\columnwidth]{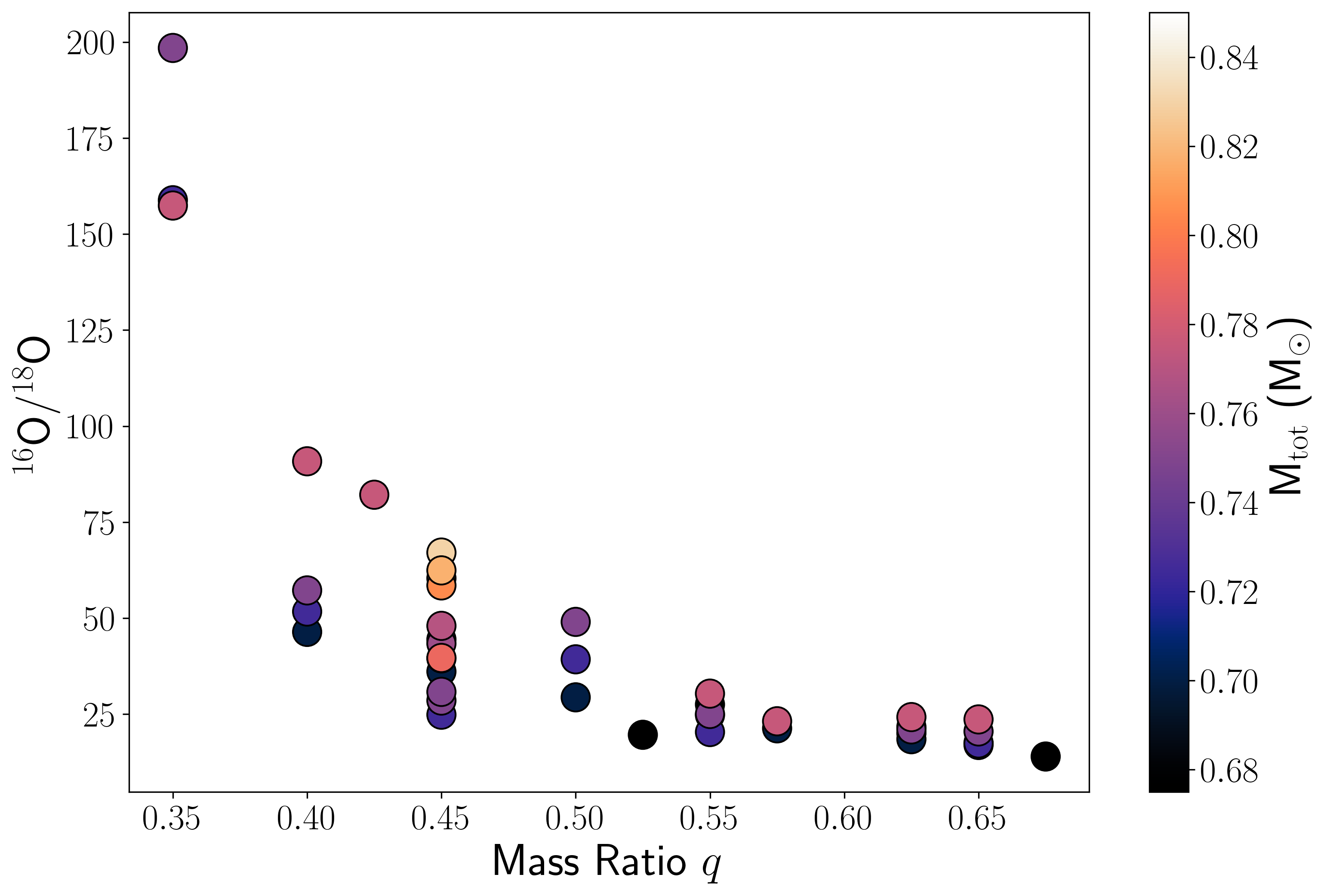}
    \caption{The oxygen isotopic ratio ${^{16}O}/{^{18}O}$ versus mass ratio $q$ and colored by \Mtot{}.
    }
    \label{fig:oxygenratio_all}
\end{figure}

We can also explore the oxygen isotopic ratio for each of our models, as the strong enhancement of $^{18}$O is characteristic of post-merger models, and part of what differentiates the HdC stars from other typical carbon stars. $^{16}$O/$^{18}$O has been studied extensively in RCBs, dLHdCs, and DY Pers, making it a useful touchstone for comparison to all three classes. In general, RCB stars are known to have $^{16}$O/$^{18}$O $\sim$ 1, but not less than 1. However, the dLHdC stars appear to have $^{16}$O/$^{18}$O $<$ 1 on average, and are therefore likely to have more surface ${18}$O than the RCBs \citep{Karambelkar2022_oxygen18}. For DY Persei, both \citet{Bhowmick2018_coolcousins} and \citet{GarciaHernandez2023_dyper} measure $^{16}$O/$^{18}$O~=~4 $\pm$ 1. \citet{Bhowmick2018_coolcousins} also studies a few DY Per ``suspect" stars for which they measure lower limits on oxygen isotopic ratios between 4 - 10. We show the oxygen isotopic ratios for our models in Figure~\ref{fig:oxygenratio_all}, which present a decrease in ratio (and thus increase in surface $^{18}$O) as one decreases \Mtot{} or increases q. The differences in surface oxygen isotopic ratio are because of small differences in the temperature and density evolution of the He-burning region. Comparing to the [F] abundances in Figure~\ref{fig:abund_F}, one can see that the trends in these two observables are indeed quite similar, as they should be.

\section{Conclusions}
\label{sec:conclusion}

Though many works have theorized that the RCB stars share an evolutionary history with the closely related classes of dLHdCs and DY Pers, and two works \citep{Tisserand2022_dLHdC,Tisserand2023_3Ddistgaia_ARXIVVERSION} discuss the likelihood of different merger populations from a population synthesis standpoint, no models have previously explored what types of DWD mergers would be needed to create such objects. In this work we aim to answer this question by varying the merger parameters of RCB models in an attempt to replicate the HRD location and abundances of these other classes of stars. To that effect we created 42 stellar engineered {\sc MESA} models with varying \Mtot{} and mass ratios, ranging from \Mtot{}~=~0.85 to 0.675 M$_\odot$ and $q$~=~0.4 to 0.675. These models show that the radius of the resultant star is a strong function of both \Mtot{} and q, with lower \Mtot{} and larger $q$ both decreasing the RCB stellar radius. This has the effect of moving the star's location in the HRD. Additionally, we find that most surface abundances (those of C, O, Ne, Na, and Mg) do not change with \Mtot{} or q, whereas initial [N] and [F] do. [N] in our models increases 0.4 to 1.0 dex over the course of the HdC phase as a function of q, due to the decay of $^{14}$C into $^{14}$N over time \citep{Munson2022_stellarengineeringpipeline}. Neither the initial nor the final [N] values agree with observations, however the final [N] values are within 0.4 dex from the observed range.
[N] is a known issue in stellar engineered HdC models, and no stellar engineered models can explain the strong enhancement seen in the RCB stars \citep{Crawford2020_RCBmodels,Munson2021_goingfrom3Dto1D,Munson2022_stellarengineeringpipeline}. 
[F] on the other hand shows strong enhancement over solar for all models. This abundance tends to decrease with decreasing \Mtot{} and increasing q, and while the majority of models do not agree with observed stars, we note that the observational range for [F] is based on only four stars that are warm enough to have measurable F lines. 
The carbon isotopic ratios of all models are consistent with observations.
Finally, we explore the oxygen isotopic ratio of our models and show that while all models have strongly enhanced surface $^{18}$O (an abundance unique to these three types of stars), the lowest \Mtot{} and largest $q$ valued models show the strongest $^{18}$O enhancement. 

With our models, we are able to make a prediction on the difference between the merger parameters for RCB stars and dLHdC stars. Recall that barring the dust production properties, which we do not probe with these models, the strongest known differences between populations of dLHdC stars and RCB stars are that dLHdCs have lower luminosities, larger \Teff{}, and increased surface $^{18}$O (lower oxygen isotopic ratio). In our set of models, those that most closely represent the dLHdC stars are those with the lowest \Mtot{} values and the largest mass ratios. In particular, our model with \Mtot{}~=~0.675 and $q$~=~0.675 has log(\Teff{})~=~3.80, log(L/L$_\odot$)~=~3.49, and $^{16}$O/$^{18}$O~=~13.99. 
While this \Teff{} and luminosity are roughly consistent with dLHdC observations, the oxygen isotopic ratio is still too large compared to observed dLHdCs. There are at least two possible ways to continue decreasing the surface $^{18}$O. The first would be to adjust the He-burning shell temperature (T$_{\rm SOF}$), as found in \citet{Crawford2020_RCBmodels} and \citet{Munson2022_stellarengineeringpipeline}. In this work, we use log(\Tsof{})~=~8.48, or roughly 301 MK, consistent with the best value from \citet{Crawford2020_RCBmodels}. However, we use the model-generation procedure from \citet{Munson2022_stellarengineeringpipeline}, which found that using a cooler \Tsof{} of 213 MK and 253 MK produced $^{16}$O/$^{18}$O~$\sim$~1, whereas \Tsof{}~=~305 MK produced a larger surface $^{16}$O/$^{18}$O~$\sim$~23. We therefore ran five extra models with \Mtot{}~=~0.675 and $q$~=~0.675 (our best models for dLHdCs) with cooler He-burning shell temperatures with a goal to further decrease the oxygen isotopic ratio. The results of this experiment are summarized in Table~\ref{tab:o-ratio}. We see that reducing the He-burning shell temperature reduces the oxygen isotopic ratio until the models at log(\Tsof{})~=~8.39 or roughly 245 MK, below which the oxygen isotopic ratio increases again as the coolest shells do not have enough He-burning to generate the necessary $^{18}$O. The lowest oxygen isotopic ratio in these five models has  $^{16}$O/$^{18}$O~=~3.76, which is much closer to the observed values for dLHdCs of $\lesssim$1. However, decreasing the He-burning shell temperature also changes the \Teff{} of the HdC phase, producing warmer HdC stars. The interplay between the merger parameters and the \Tsof{} requires a delicate balance to replicate observations, which we leave to further works.
However, it may also be possible to further reduce \Mtot{} and increase $q$ in attempt to move even more towards the dLHdC observations, however we ran into numerical issues with {\sc MESA} in this regime.
We additionally know that there is a lower limit on \Mtot{} for which the accretor star of the DWD binary can no longer be a CO-WD or a ``hybrid"-WD as in the population synthesis models of \citet{Ruiter2019_popsynth} and therefore our modelling process would not be appropriate for such a merger. The merger of two He-WDs has been studied in \citet{Schwab2018_hehemergers}, and they do not resemble dLHdC stars. 
We therefore also leave the study of lower \Mtot{} merger products to future works.

\begin{table}
    \centering
    \caption{Five cool models at \Mtot{}~=~0.675 and $q$~=~0.675 to test $^{16}$O/$^{18}$O}
    \label{tab:o-ratio}
    \begin{tabular}{cccc}
        log(\Tsof{}) & $^{16}$O/$^{18}$O & log(T$_{\text{eff,HdC}}$) & log(L$_{\rm HdC}$) \\
        \hline
        8.43 & 5.29 & 3.84 & 3.48 \\
        8.39 & 3.76 & 3.86 & 3.48 \\
        8.34 & 73.08 & 3.92 & 3.48 \\
        8.32 & 1.85$\times$10$^{5}$ & 3.93 & 3.49 \\
        8.16 & 6.04$\times$10$^{7}$ & 3.92 & 3.57 \\
    \end{tabular}
\end{table}


Contrary to the dLHdCs, none of our models can explain the formation of the DY Per type variables. DY Per stars, while also having different dust properties from RCB stars, have lower luminosities than RCB stars but are generally much cooler, with DY Per having \Teff{} $\sim$ 3500 K \citep{Keenan1997_DYPerseiTemperature}. This cool temperature makes it a particularly challenging star for abundance studies, with \citet{GarciaHernandez2023_dyper} referring to it as ``a spectroscopist's nightmare." Not much work has been done on their surface abundances in the optical, however both \citet{Bhowmick2018_coolcousins} and \citet{GarciaHernandez2023_dyper} show that DY Per's infrared spectra show enhanced $^{18}$O on the same level as the RCB stars. \citet{GarciaHernandez2023_dyper} was unable to definitively confirm DY Per's [H] abundance, but finds evidence that the star may be H-deficient, although likely not as H-deficient as RCBs. An important detail is that \citet{Zacs2007_dyper}'s optical models of DY Per suggest that the star is best fit with solar metallicity. \citet{Crawford2020_RCBmodels} previously showed that solar metallicity models tend to be cooler than subsolar metallicity models. This implies that perhaps if we were to increase the initial metallicity of our merger models that we could potentially move them towards the HRD location of the DY Per stars. Additionally, \citet{Schwab2019_2DintoMesaSimulations} showed that by using a more appropriate opacity table 
for carbon-rich atmospheres or by increasing the envelope carbon fraction to directly increase the envelope opacity, one
could cool RCB models made via modified He-star evolution, which could once again move us towards the location of DY Pers. We leave explorations of these topics to later works.

This work has focused predominantly on the HdC stars found in this specific region of the HRD, but in doing so we miss discussion of yet another class of star which is hypothesized to share an evolutionary history with the RCBs called the Extreme Helium (EHe) Stars. These are rare (24 members) low-mass supergiant stars with much higher temperatures than RCB stars, ranging from $\sim$9000 - 35000K \citep{Pandey2006_ehetemps}. There are additionally five known RCB stars in this temperature range \citep{Clayton2012_HdCMasses}, and the only known difference between these and the EHe stars is that the RCB stars still exhibit dust declines. One current theory is that when the RCB stars evolve further, they move leftwards in the HRD, in which they pass through the location where they would appear as EHe stars. There are a few discrepancies in this theory, however, as the EHe stars tend to show slightly different abundances than RCBs \citep{Jeffery2011_abundances}. 

In this work we have shown that DWD merger parameters can affect the \Teff{} and luminosity of an RCB-like model, as well as its [N], [F], and oxygen isotopic ratio. We find strong evidence to support the claim that dLHdC stars are formed via a slightly different merger population than the RCB stars, where the dLHdCs will have lower \Mtot{} and higher mass ratios. Further work must be done to understand how these differences would result in a difference in dust production properties. We are not able to explain the existence of DY Per type variables with our models, however there are two possible changes that could rectify this, which are the metallicity and the adopted opacity tables. Additionally, our models are limited in that they all begin from identical initial abundances, and do not take into account structural differences in the CO-WD core as it changes mass. We hope to address more of these details in future models as the stellar engineering process and {\sc MESA} itself continues to evolve and improve over time.


\section*{Acknowledgements}

We would like to thank Bradley Munson and Yaguang Li for their correspondence and advice throughout the drafting of this paper.
We would also like to thank Sunny Wong and Simon Jeffery for discussions on the results of this paper that led us to a deeper understanding of the model physics. We finally thank the anonymous reviewer for their time and their helpful comments.

This research was undertaken with the assistance of resources and services from the National Computational Infrastructure (NCI), which is supported by the Australian Government. C.C. gratefully acknowledges support from the Australian Research Council through Discovery Project DP210103119. M.G.P. acknowledges support from the Professor Harry Messel Research Fellowship in Physics Endowment, at the University of Sydney.

This work made extensive use of the stellar evolution code Modules for Experiments in Stellar Astrophysics
\citep[{\sc MESA}][]{Paxton2011, Paxton2013, Paxton2015, Paxton2018, Paxton2019, Jermyn2023}. The {\sc MESA} EOS is a blend of the OPAL \citep{Rogers2002}, SCVH
\citep{Saumon1995}, FreeEOS \citep{Irwin2004}, HELM \citep{Timmes2000},
PC \citep{Potekhin2010}, and Skye \citep{Jermyn2021} EOSes.
Radiative opacities are primarily from OPAL \citep{Iglesias1993,
Iglesias1996}, with low-temperature data from \citet{Ferguson2005}
and the high-temperature, Compton-scattering dominated regime by
\citet{Poutanen2017}.  Electron conduction opacities are from
\citet{Cassisi2007} and \citet{Blouin2020}.
Nuclear reaction rates are from JINA REACLIB \citep{Cyburt2010}, NACRE \citep{Angulo1999} and
additional tabulated weak reaction rates \citet{Fuller1985, Oda1994,
Langanke2000}.  Screening is included via the prescription of \citet{Chugunov2007}.
Thermal neutrino loss rates are from \citet{Itoh1996}.

This work made use of several publicly available {\tt python} packages: {\tt astropy} \citep{astropy:2013,astropy:2018}, 
{\tt matplotlib} \citep{matplotlib2007}, 
{\tt numpy} \citep{numpy2020}, and 
{\tt scipy} \citep{scipy2020}.

\section*{Data Availability}

The {\sc MESA} models generated in this work are available upon reasonable request to the corresponding author.



\bibliographystyle{mnras}
\bibliography{dyper,mesa} 

\begin{thebibliography}{}
\makeatletter
\relax
\def\mn@urlcharsother{\let\do\@makeother \do\$\do\&\do\#\do\^\do\_\do\%\do\~}
\def\mn@doi{\begingroup\mn@urlcharsother \@ifnextchar [ {\mn@doi@} {\mn@doi@[]}}
\def\mn@doi@[#1]#2{\def\@tempa{#1}\ifx\@tempa\@empty \href {http://dx.doi.org/#2} {doi:#2}\else \href {http://dx.doi.org/#2} {#1}\fi \endgroup}
\def\mn@eprint#1#2{\mn@eprint@#1:#2::\@nil}
\def\mn@eprint@arXiv#1{\href {http://arxiv.org/abs/#1} {{\tt arXiv:#1}}}
\def\mn@eprint@dblp#1{\href {http://dblp.uni-trier.de/rec/bibtex/#1.xml} {dblp:#1}}
\def\mn@eprint@#1:#2:#3:#4\@nil{\def\@tempa {#1}\def\@tempb {#2}\def\@tempc {#3}\ifx \@tempc \@empty \let \@tempc \@tempb \let \@tempb \@tempa \fi \ifx \@tempb \@empty \def\@tempb {arXiv}\fi \@ifundefined {mn@eprint@\@tempb}{\@tempb:\@tempc}{\expandafter \expandafter \csname mn@eprint@\@tempb\endcsname \expandafter{\@tempc}}}

\bibitem[\protect\citeauthoryear{{Alcock} et~al.,}{{Alcock} et~al.}{2001}]{Alcock2001_newmachorcb_dyper}
{Alcock} C.,  et~al., 2001, \mn@doi [\apj] {10.1086/321369}, \href {https://ui.adsabs.harvard.edu/abs/2001ApJ...554..298A} {554, 298}

\bibitem[\protect\citeauthoryear{{Alksnis}}{{Alksnis}}{1994}]{Alksnis1994_dyper}
{Alksnis} A.,  1994, \mn@doi [Baltic Astronomy] {10.1515/astro-1994-0406}, \href {https://ui.adsabs.harvard.edu/abs/1994BaltA...3..410A} {3, 410}

\bibitem[\protect\citeauthoryear{{Angulo} et~al.,}{{Angulo} et~al.}{1999}]{Angulo1999}
{Angulo} C.,  et~al., 1999, \mn@doi [\nphysa] {10.1016/S0375-9474(99)00030-5}, \href {https://ui.adsabs.harvard.edu/abs/1999NuPhA.656....3A} {656, 3}

\bibitem[\protect\citeauthoryear{{Asplund}, {Gustafsson}, {Lambert}  \& {Rao}}{{Asplund} et~al.}{2000}]{Asplund2000_rcbabundances}
{Asplund} M.,  {Gustafsson} B.,  {Lambert} D.~L.,   {Rao} N.~K.,  2000, \aap, \href {https://ui.adsabs.harvard.edu/abs/2000A&A...353..287A} {353, 287}

\bibitem[\protect\citeauthoryear{{Astropy Collaboration}}{{Astropy Collaboration}}{2013}]{astropy:2013}
{Astropy Collaboration} 2013, \mn@doi [\aap] {10.1051/0004-6361/201322068}, \href {http://adsabs.harvard.edu/abs/2013A%26A...558A..33A} {558, A33}

\bibitem[\protect\citeauthoryear{{Astropy Collaboration}}{{Astropy Collaboration}}{2018}]{astropy:2018}
{Astropy Collaboration} 2018, \mn@doi [\aj] {10.3847/1538-3881/aabc4f}, \href {https://ui.adsabs.harvard.edu/abs/2018AJ....156..123A} {156, 123}

\bibitem[\protect\citeauthoryear{{Barnbaum}, {Stone}  \& {Keenan}}{{Barnbaum} et~al.}{1996}]{Barnbaum1996_carbonstaratlas}
{Barnbaum} C.,  {Stone} R. P.~S.,   {Keenan} P.~C.,  1996, \mn@doi [\apjs] {10.1086/192323}, \href {https://ui.adsabs.harvard.edu/abs/1996ApJS..105..419B} {105, 419}

\bibitem[\protect\citeauthoryear{{Bhowmick}, {Pandey}, {Joshi}  \& {Ashok}}{{Bhowmick} et~al.}{2018}]{Bhowmick2018_coolcousins}
{Bhowmick} A.,  {Pandey} G.,  {Joshi} V.,   {Ashok} N.~M.,  2018, \mn@doi [\apj] {10.3847/1538-4357/aaaae4}, \href {https://ui.adsabs.harvard.edu/abs/2018ApJ...854..140B} {854, 140}

\bibitem[\protect\citeauthoryear{{Bhowmick}, {Pandey}  \& {Lambert}}{{Bhowmick} et~al.}{2020}]{Bhowmick2020_FluorineinEHeStars}
{Bhowmick} A.,  {Pandey} G.,   {Lambert} D.~L.,  2020, \mn@doi [\apj] {10.3847/1538-4357/ab6e6d}, \href {https://ui.adsabs.harvard.edu/abs/2020ApJ...891...40B} {891, 40}

\bibitem[\protect\citeauthoryear{{Bloecker}}{{Bloecker}}{1995}]{Bloecker1995_massloss}
{Bloecker} T.,  1995, \aap, \href {https://ui.adsabs.harvard.edu/abs/1995A&A...297..727B} {297, 727}

\bibitem[\protect\citeauthoryear{{Blouin}, {Shaffer}, {Saumon}  \& {Starrett}}{{Blouin} et~al.}{2020}]{Blouin2020}
{Blouin} S.,  {Shaffer} N.~R.,  {Saumon} D.,   {Starrett} C.~E.,  2020, \mn@doi [\apj] {10.3847/1538-4357/ab9e75}, \href {https://ui.adsabs.harvard.edu/abs/2020ApJ...899...46B} {899, 46}

\bibitem[\protect\citeauthoryear{{Cassisi}, {Potekhin}, {Pietrinferni}, {Catelan}  \& {Salaris}}{{Cassisi} et~al.}{2007}]{Cassisi2007}
{Cassisi} S.,  {Potekhin} A.~Y.,  {Pietrinferni} A.,  {Catelan} M.,   {Salaris} M.,  2007, \mn@doi [\apj] {10.1086/516819}, \href {https://ui.adsabs.harvard.edu/abs/2007ApJ...661.1094C} {661, 1094}

\bibitem[\protect\citeauthoryear{{Chugunov}, {Dewitt}  \& {Yakovlev}}{{Chugunov} et~al.}{2007}]{Chugunov2007}
{Chugunov} A.~I.,  {Dewitt} H.~E.,   {Yakovlev} D.~G.,  2007, \mn@doi [\prd] {10.1103/PhysRevD.76.025028}, \href {https://ui.adsabs.harvard.edu/abs/2007PhRvD..76b5028C} {76, 025028}

\bibitem[\protect\citeauthoryear{{Clayton}}{{Clayton}}{1996}]{Clayton1996_HdCcomposition}
{Clayton} G.~C.,  1996, \mn@doi [\pasp] {10.1086/133715}, \href {https://ui.adsabs.harvard.edu/abs/1996PASP..108..225C} {108, 225}

\bibitem[\protect\citeauthoryear{{Clayton}}{{Clayton}}{2012}]{Clayton2012_HdCMasses}
{Clayton} G.~C.,  2012, in American Astronomical Society Meeting Abstracts \#219. p. 152.01

\bibitem[\protect\citeauthoryear{{Clayton}, {Geballe}, {Herwig}, {Fryer}  \& {Asplund}}{{Clayton} et~al.}{2007}]{Clayton2007_Excess018inRCBsandHDCsbecauseWDM}
{Clayton} G.~C.,  {Geballe} T.~R.,  {Herwig} F.,  {Fryer} C.,   {Asplund} M.,  2007, \mn@doi [\apj] {10.1086/518307}, \href {https://ui.adsabs.harvard.edu/abs/2007ApJ...662.1220C} {662, 1220}

\bibitem[\protect\citeauthoryear{{Crawford}, {Clayton}, {Munson}, {Chatzopoulos}  \& {Frank}}{{Crawford} et~al.}{2020}]{Crawford2020_RCBmodels}
{Crawford} C.~L.,  {Clayton} G.~C.,  {Munson} B.,  {Chatzopoulos} E.,   {Frank} J.,  2020, \mn@doi [\mnras] {10.1093/mnras/staa2526}, \href {https://ui.adsabs.harvard.edu/abs/2020MNRAS.498.2912C} {498, 2912}

\bibitem[\protect\citeauthoryear{{Crawford} et~al.,}{{Crawford} et~al.}{2023}]{Crawford2023_hdcclassification}
{Crawford} C.~L.,  et~al., 2023, \mn@doi [\mnras] {10.1093/mnras/stad324}, \href {https://ui.adsabs.harvard.edu/abs/2023MNRAS.521.1674C} {521, 1674}

\bibitem[\protect\citeauthoryear{{Cyburt} et~al.,}{{Cyburt} et~al.}{2010}]{Cyburt2010}
{Cyburt} R.~H.,  et~al., 2010, \mn@doi [\apjs] {10.1088/0067-0049/189/1/240}, \href {https://ui.adsabs.harvard.edu/abs/2010ApJS..189..240C} {189, 240}

\bibitem[\protect\citeauthoryear{{Denissenkov} et~al.,}{{Denissenkov} et~al.}{2018}]{Denissenkov2018_iprocess}
{Denissenkov} P.,  et~al., 2018, \mn@doi [Journal of Physics G Nuclear Physics] {10.1088/1361-6471/aabb6e}, \href {https://ui.adsabs.harvard.edu/abs/2018JPhG...45e5203D} {45, 055203}

\bibitem[\protect\citeauthoryear{{Ferguson}, {Alexander}, {Allard}, {Barman}, {Bodnarik}, {Hauschildt}, {Heffner-Wong}  \& {Tamanai}}{{Ferguson} et~al.}{2005}]{Ferguson2005}
{Ferguson} J.~W.,  {Alexander} D.~R.,  {Allard} F.,  {Barman} T.,  {Bodnarik} J.~G.,  {Hauschildt} P.~H.,  {Heffner-Wong} A.,   {Tamanai} A.,  2005, \mn@doi [\apj] {10.1086/428642}, \href {https://ui.adsabs.harvard.edu/abs/2005ApJ...623..585F} {623, 585}

\bibitem[\protect\citeauthoryear{{Fuller}, {Fowler}  \& {Newman}}{{Fuller} et~al.}{1985}]{Fuller1985}
{Fuller} G.~M.,  {Fowler} W.~A.,   {Newman} M.~J.,  1985, \mn@doi [\apj] {10.1086/163208}, \href {https://ui.adsabs.harvard.edu/abs/1985ApJ...293....1F} {293, 1}

\bibitem[\protect\citeauthoryear{{Garc{\'{\i}}a-Hern{\'a}ndez}, {Lambert}, {Kameswara Rao}, {Hinkle}  \& {Eriksson}}{{Garc{\'{\i}}a-Hern{\'a}ndez} et~al.}{2010}]{GarciaHernandez2010_oxygen18}
{Garc{\'{\i}}a-Hern{\'a}ndez} D.~A.,  {Lambert} D.~L.,  {Kameswara Rao} N.,  {Hinkle} K.~H.,   {Eriksson} K.,  2010, \apj, 714, 144

\bibitem[\protect\citeauthoryear{{Garc{\'\i}a-Hern{\'a}ndez}, {Rao}  \& {Lambert}}{{Garc{\'\i}a-Hern{\'a}ndez} et~al.}{2013}]{GarciaHernandez2013_dyperPAH}
{Garc{\'\i}a-Hern{\'a}ndez} D.~A.,  {Rao} N.~K.,   {Lambert} D.~L.,  2013, \mn@doi [\apj] {10.1088/0004-637X/773/2/107}, \href {https://ui.adsabs.harvard.edu/abs/2013ApJ...773..107G} {773, 107}

\bibitem[\protect\citeauthoryear{{Garc{\'\i}a-Hern{\'a}ndez}, {Rao}, {Lambert}, {Eriksson}, {Reddy}  \& {Masseron}}{{Garc{\'\i}a-Hern{\'a}ndez} et~al.}{2023}]{GarciaHernandez2023_dyper}
{Garc{\'\i}a-Hern{\'a}ndez} D.~A.,  {Rao} N.~K.,  {Lambert} D.~L.,  {Eriksson} K.,  {Reddy} A.~B.~S.,   {Masseron} T.,  2023, \mn@doi [\apj] {10.3847/1538-4357/acc574}, \href {https://ui.adsabs.harvard.edu/abs/2023ApJ...948...15G} {948, 15}

\bibitem[\protect\citeauthoryear{{Gautschy}}{{Gautschy}}{2023}]{Gautschy2023_rcbpulsation_ARXIVVERSION}
{Gautschy} A.,  2023, \mn@doi [arXiv e-prints] {10.48550/arXiv.2312.14693}, \href {https://ui.adsabs.harvard.edu/abs/2023arXiv231214693G} {p. arXiv:2312.14693}

\bibitem[\protect\citeauthoryear{{Goldstein} \& {Townsend}}{{Goldstein} \& {Townsend}}{2020}]{Goldstein2020_gyre}
{Goldstein} J.,  {Townsend} R.~H.~D.,  2020, \mn@doi [\apj] {10.3847/1538-4357/aba748}, \href {https://ui.adsabs.harvard.edu/abs/2020ApJ...899..116G} {899, 116}

\bibitem[\protect\citeauthoryear{{Grevesse} \& {Sauval}}{{Grevesse} \& {Sauval}}{1998}]{GS98_metalmixture}
{Grevesse} N.,  {Sauval} A.~J.,  1998, \mn@doi [\ssr] {10.1023/A:1005161325181}, \href {https://ui.adsabs.harvard.edu/abs/1998SSRv...85..161G} {85, 161}

\bibitem[\protect\citeauthoryear{Harris et~al.,}{Harris et~al.}{2020}]{numpy2020}
Harris C.~R.,  et~al., 2020, \mn@doi [Nature] {10.1038/s41586-020-2649-2}, 585, 357

\bibitem[\protect\citeauthoryear{{Herwig} et~al.,}{{Herwig} et~al.}{2008}]{Herwig2008_nugrid}
{Herwig} F.,  et~al., 2008, in Nuclei in the Cosmos (NIC X). p.~E23 (\mn@eprint {arXiv} {0811.4653}), \mn@doi{10.22323/1.053.0023}

\bibitem[\protect\citeauthoryear{{Howell}, {Rector}  \& {Walter}}{{Howell} et~al.}{2013}]{Howell2013_dustmagnitudedrop8}
{Howell} S.~B.,  {Rector} T.~A.,   {Walter} D.,  2013, \mn@doi [\pasp] {10.1086/672163}, \href {https://ui.adsabs.harvard.edu/abs/2013PASP..125..879H} {125, 879}

\bibitem[\protect\citeauthoryear{Hunter}{Hunter}{2007}]{matplotlib2007}
Hunter J.~D.,  2007, Computing in Science \& Engineering, 9, 90

\bibitem[\protect\citeauthoryear{{Iben} \& {Tutukov}}{{Iben} \& {Tutukov}}{1984}]{1984ApJS...54..335I}
{Iben} I. J.,  {Tutukov} A.~V.,  1984, \mn@doi [\apjs] {10.1086/190932}, \href {https://ui.adsabs.harvard.edu/abs/1984ApJS...54..335I} {54, 335}

\bibitem[\protect\citeauthoryear{{Iglesias} \& {Rogers}}{{Iglesias} \& {Rogers}}{1993}]{Iglesias1993}
{Iglesias} C.~A.,  {Rogers} F.~J.,  1993, \mn@doi [\apj] {10.1086/172958}, \href {https://ui.adsabs.harvard.edu/abs/1993ApJ...412..752I} {412, 752}

\bibitem[\protect\citeauthoryear{{Iglesias} \& {Rogers}}{{Iglesias} \& {Rogers}}{1996}]{Iglesias1996}
{Iglesias} C.~A.,  {Rogers} F.~J.,  1996, \mn@doi [\apj] {10.1086/177381}, \href {https://ui.adsabs.harvard.edu/abs/1996ApJ...464..943I} {464, 943}

\bibitem[\protect\citeauthoryear{{Irwin}}{{Irwin}}{2004}]{Irwin2004}
{Irwin} A.~W.,  2004, The FreeEOS Code for Calculating the Equation of State for Stellar Interiors, \url {http://freeeos.sourceforge.net/}

\bibitem[\protect\citeauthoryear{{Itoh}, {Hayashi}, {Nishikawa}  \& {Kohyama}}{{Itoh} et~al.}{1996}]{Itoh1996}
{Itoh} N.,  {Hayashi} H.,  {Nishikawa} A.,   {Kohyama} Y.,  1996, \mn@doi [\apjs] {10.1086/192264}, \href {https://ui.adsabs.harvard.edu/abs/1996ApJS..102..411I} {102, 411}

\bibitem[\protect\citeauthoryear{{Jeffery}}{{Jeffery}}{1988}]{Jeffery1988_coremassluminosity}
{Jeffery} C.~S.,  1988, \mn@doi [\mnras] {10.1093/mnras/235.4.1287}, \href {https://ui.adsabs.harvard.edu/abs/1988MNRAS.235.1287J} {235, 1287}

\bibitem[\protect\citeauthoryear{{Jeffery}, {Karakas}  \& {Saio}}{{Jeffery} et~al.}{2011}]{Jeffery2011_abundances}
{Jeffery} C.~S.,  {Karakas} A.~I.,   {Saio} H.,  2011, \mn@doi [\mnras] {10.1111/j.1365-2966.2011.18667.x}, \href {https://ui.adsabs.harvard.edu/abs/2011MNRAS.414.3599J} {414, 3599}

\bibitem[\protect\citeauthoryear{{Jermyn}, {Schwab}, {Bauer}, {Timmes}  \& {Potekhin}}{{Jermyn} et~al.}{2021}]{Jermyn2021}
{Jermyn} A.~S.,  {Schwab} J.,  {Bauer} E.,  {Timmes} F.~X.,   {Potekhin} A.~Y.,  2021, \mn@doi [\apj] {10.3847/1538-4357/abf48e}, \href {https://ui.adsabs.harvard.edu/abs/2021ApJ...913...72J} {913, 72}

\bibitem[\protect\citeauthoryear{{Jermyn} et~al.,}{{Jermyn} et~al.}{2023}]{Jermyn2023}
{Jermyn} A.~S.,  et~al., 2023, \mn@doi [\apjs] {10.3847/1538-4365/acae8d}, \href {https://ui.adsabs.harvard.edu/abs/2023ApJS..265...15J} {265, 15}

\bibitem[\protect\citeauthoryear{{Karambelkar}, {Kasliwal}, {Tisserand}, {Clayton}, {Crawford}, {Anand}, {Geballe}  \& {Montiel}}{{Karambelkar} et~al.}{2022}]{Karambelkar2022_oxygen18}
{Karambelkar} V.,  {Kasliwal} M.~M.,  {Tisserand} P.,  {Clayton} G.~C.,  {Crawford} C.~L.,  {Anand} S.~G.,  {Geballe} T.~R.,   {Montiel} E.,  2022, \mn@doi [\aap] {10.1051/0004-6361/202142918}, \href {https://ui.adsabs.harvard.edu/abs/2022A&A...667A..84K} {667, A84}

\bibitem[\protect\citeauthoryear{{Keenan} \& {Barnbaum}}{{Keenan} \& {Barnbaum}}{1997}]{Keenan1997_DYPerseiTemperature}
{Keenan} P.~C.,  {Barnbaum} C.,  1997, \mn@doi [\pasp] {10.1086/133968}, \href {https://ui.adsabs.harvard.edu/abs/1997PASP..109..969K} {109, 969}

\bibitem[\protect\citeauthoryear{{Kipper}}{{Kipper}}{2002}]{Kipper2002_dlhdcabunds_hmlib}
{Kipper} T.,  2002, Baltic Astronomy, \href {https://ui.adsabs.harvard.edu/abs/2002BaltA..11..249K} {11, 249}

\bibitem[\protect\citeauthoryear{{Langanke} \& {Mart{\'{\i}}nez-Pinedo}}{{Langanke} \& {Mart{\'{\i}}nez-Pinedo}}{2000}]{Langanke2000}
{Langanke} K.,  {Mart{\'{\i}}nez-Pinedo} G.,  2000, \mn@doi [Nuclear Physics A] {10.1016/S0375-9474(00)00131-7}, \href {https://ui.adsabs.harvard.edu/abs/2000NuPhA.673..481L} {673, 481}

\bibitem[\protect\citeauthoryear{{Lauer}, {Chatzopoulos}, {Clayton}, {Frank}  \& {Marcello}}{{Lauer} et~al.}{2019}]{Lauer2019_OriginalStellarEngineering}
{Lauer} A.,  {Chatzopoulos} E.,  {Clayton} G.~C.,  {Frank} J.,   {Marcello} D.~C.,  2019, \mn@doi [\mnras] {10.1093/mnras/stz1732}, \href {https://ui.adsabs.harvard.edu/abs/2019MNRAS.488..438L} {488, 438}

\bibitem[\protect\citeauthoryear{Lodders}{Lodders}{2003}]{Lodders2003_SolarAbundances}
Lodders K.,  2003, \mn@doi [The Astrophysical Journal] {10.1086/375492}, 591, 1220

\bibitem[\protect\citeauthoryear{{Longland}, {Lor{\'e}n-Aguilar}, {Jos{\'e}}, {Garc{\'\i}a-Berro}, {Althaus}  \& {Isern}}{{Longland} et~al.}{2011}]{Longland2011_hydrorcbmodel}
{Longland} R.,  {Lor{\'e}n-Aguilar} P.,  {Jos{\'e}} J.,  {Garc{\'\i}a-Berro} E.,  {Althaus} L.~G.,   {Isern} J.,  2011, \mn@doi [\apjl] {10.1088/2041-8205/737/2/L34}, \href {https://ui.adsabs.harvard.edu/abs/2011ApJ...737L..34L} {737, L34}

\bibitem[\protect\citeauthoryear{{Marigo} \& {Aringer}}{{Marigo} \& {Aringer}}{2009}]{MarigoAringer2009_aesopus}
{Marigo} P.,  {Aringer} B.,  2009, \mn@doi [\aap] {10.1051/0004-6361/200912598}, \href {https://ui.adsabs.harvard.edu/abs/2009A&A...508.1539M} {508, 1539}

\bibitem[\protect\citeauthoryear{{Marigo}, {Aringer}, {Girardi}  \& {Bressan}}{{Marigo} et~al.}{2022}]{Marigo2022_aesopus2}
{Marigo} P.,  {Aringer} B.,  {Girardi} L.,   {Bressan} A.,  2022, \mn@doi [\apj] {10.3847/1538-4357/ac9b40}, \href {https://ui.adsabs.harvard.edu/abs/2022ApJ...940..129M} {940, 129}

\bibitem[\protect\citeauthoryear{{Menon}, {Herwig}, {Denissenkov}, {Clayton}, {Staff}, {Pignatari}  \& {Paxton}}{{Menon} et~al.}{2013}]{Menon2013_ModifiedAGBevolutionOlder}
{Menon} A.,  {Herwig} F.,  {Denissenkov} P.~A.,  {Clayton} G.~C.,  {Staff} J.,  {Pignatari} M.,   {Paxton} B.,  2013, \mn@doi [\apj] {10.1088/0004-637X/772/1/59}, \href {https://ui.adsabs.harvard.edu/abs/2013ApJ...772...59M} {772, 59}

\bibitem[\protect\citeauthoryear{{Menon}, {Karakas}, {Lugaro}, {Doherty}  \& {Ritter}}{{Menon} et~al.}{2019}]{Menon2019_AGBevolutionofRCBs}
{Menon} A.,  {Karakas} A.~I.,  {Lugaro} M.,  {Doherty} C.~L.,   {Ritter} C.,  2019, \mn@doi [\mnras] {10.1093/mnras/sty2606}, \href {https://ui.adsabs.harvard.edu/abs/2019MNRAS.482.2320M} {482, 2320}

\bibitem[\protect\citeauthoryear{{Munson}, {Chatzopoulos}, {Frank}, {Clayton}, {Crawford}, {Denissenkov}  \& {Herwig}}{{Munson} et~al.}{2021}]{Munson2021_goingfrom3Dto1D}
{Munson} B.,  {Chatzopoulos} E.,  {Frank} J.,  {Clayton} G.~C.,  {Crawford} C.~L.,  {Denissenkov} P.~A.,   {Herwig} F.,  2021, \mn@doi [\apj] {10.3847/1538-4357/abeb6c}, \href {https://ui.adsabs.harvard.edu/abs/2021ApJ...911..103M} {911, 103}

\bibitem[\protect\citeauthoryear{{Munson}, {Chatzopoulos}  \& {Denissenkov}}{{Munson} et~al.}{2022}]{Munson2022_stellarengineeringpipeline}
{Munson} B.,  {Chatzopoulos} E.,   {Denissenkov} P.~A.,  2022, \mn@doi [\apj] {10.3847/1538-4357/ac9476}, \href {https://ui.adsabs.harvard.edu/abs/2022ApJ...939...45M} {939, 45}

\bibitem[\protect\citeauthoryear{{Oda}, {Hino}, {Muto}, {Takahara}  \& {Sato}}{{Oda} et~al.}{1994}]{Oda1994}
{Oda} T.,  {Hino} M.,  {Muto} K.,  {Takahara} M.,   {Sato} K.,  1994, \mn@doi [Atomic Data and Nuclear Data Tables] {10.1006/adnd.1994.1007}, \href {https://ui.adsabs.harvard.edu/abs/1994ADNDT..56..231O} {56, 231}

\bibitem[\protect\citeauthoryear{{Paczy{\'n}ski}}{{Paczy{\'n}ski}}{1970}]{Paczynski1970_coremassluminosity}
{Paczy{\'n}ski} B.,  1970, \actaa, \href {https://ui.adsabs.harvard.edu/abs/1970AcA....20...47P} {20, 47}

\bibitem[\protect\citeauthoryear{{Pandey}, {Lambert}, {Jeffery}  \& {Rao}}{{Pandey} et~al.}{2006}]{Pandey2006_ehetemps}
{Pandey} G.,  {Lambert} D.~L.,  {Jeffery} C.~S.,   {Rao} N.~K.,  2006, \mn@doi [\apj] {10.1086/498674}, \href {https://ui.adsabs.harvard.edu/abs/2006ApJ...638..454P} {638, 454}

\bibitem[\protect\citeauthoryear{{Paxton}, {Bildsten}, {Dotter}, {Herwig}, {Lesaffre}  \& {Timmes}}{{Paxton} et~al.}{2011}]{Paxton2011}
{Paxton} B.,  {Bildsten} L.,  {Dotter} A.,  {Herwig} F.,  {Lesaffre} P.,   {Timmes} F.,  2011, \mn@doi [\apjs] {10.1088/0067-0049/192/1/3}, \href {https://ui.adsabs.harvard.edu/abs/2011ApJS..192....3P} {192, 3}

\bibitem[\protect\citeauthoryear{{Paxton} et~al.,}{{Paxton} et~al.}{2013}]{Paxton2013}
{Paxton} B.,  et~al., 2013, \mn@doi [\apjs] {10.1088/0067-0049/208/1/4}, \href {https://ui.adsabs.harvard.edu/abs/2013ApJS..208....4P} {208, 4}

\bibitem[\protect\citeauthoryear{{Paxton} et~al.,}{{Paxton} et~al.}{2015}]{Paxton2015}
{Paxton} B.,  et~al., 2015, \mn@doi [\apjs] {10.1088/0067-0049/220/1/15}, \href {https://ui.adsabs.harvard.edu/abs/2015ApJS..220...15P} {220, 15}

\bibitem[\protect\citeauthoryear{{Paxton} et~al.,}{{Paxton} et~al.}{2018}]{Paxton2018}
{Paxton} B.,  et~al., 2018, \mn@doi [\apjs] {10.3847/1538-4365/aaa5a8}, \href {https://ui.adsabs.harvard.edu/abs/2018ApJS..234...34P} {234, 34}

\bibitem[\protect\citeauthoryear{{Paxton} et~al.,}{{Paxton} et~al.}{2019}]{Paxton2019}
{Paxton} B.,  et~al., 2019, \mn@doi [\apjs] {10.3847/1538-4365/ab2241}, \href {https://ui.adsabs.harvard.edu/abs/2019ApJS..243...10P} {243, 10}

\bibitem[\protect\citeauthoryear{{Pignatari} \& {Herwig}}{{Pignatari} \& {Herwig}}{2012}]{Pignatari2012_nugrid}
{Pignatari} M.,  {Herwig} F.,  2012, \mn@doi [Nuclear Physics News] {10.1080/10619127.2012.710142}, \href {https://ui.adsabs.harvard.edu/abs/2012NPNew..22...18P} {22, 18}

\bibitem[\protect\citeauthoryear{{Potekhin} \& {Chabrier}}{{Potekhin} \& {Chabrier}}{2010}]{Potekhin2010}
{Potekhin} A.~Y.,  {Chabrier} G.,  2010, \mn@doi [Contributions to Plasma Physics] {10.1002/ctpp.201010017}, \href {https://ui.adsabs.harvard.edu/abs/2010CoPP...50...82P} {50, 82}

\bibitem[\protect\citeauthoryear{{Poutanen}}{{Poutanen}}{2017}]{Poutanen2017}
{Poutanen} J.,  2017, \mn@doi [\apj] {10.3847/1538-4357/835/2/119}, \href {https://ui.adsabs.harvard.edu/abs/2017ApJ...835..119P} {835, 119}

\bibitem[\protect\citeauthoryear{{Rogers} \& {Nayfonov}}{{Rogers} \& {Nayfonov}}{2002}]{Rogers2002}
{Rogers} F.~J.,  {Nayfonov} A.,  2002, \mn@doi [\apj] {10.1086/341894}, \href {https://ui.adsabs.harvard.edu/abs/2002ApJ...576.1064R} {576, 1064}

\bibitem[\protect\citeauthoryear{{Ruiter}, {Ferrario}, {Belczynski}, {Seitenzahl}, {Crocker}  \& {Karakas}}{{Ruiter} et~al.}{2019}]{Ruiter2019_popsynth}
{Ruiter} A.~J.,  {Ferrario} L.,  {Belczynski} K.,  {Seitenzahl} I.~R.,  {Crocker} R.~M.,   {Karakas} A.~I.,  2019, \mn@doi [\mnras] {10.1093/mnras/stz001}, \href {https://ui.adsabs.harvard.edu/abs/2019MNRAS.484..698R} {484, 698}

\bibitem[\protect\citeauthoryear{{Saio}}{{Saio}}{1988}]{Saio1988_coremassluminosity}
{Saio} H.,  1988, \mn@doi [\mnras] {10.1093/mnras/235.1.203}, \href {https://ui.adsabs.harvard.edu/abs/1988MNRAS.235..203S} {235, 203}

\bibitem[\protect\citeauthoryear{{Saio} \& {Jeffery}}{{Saio} \& {Jeffery}}{2000}]{SaioJeffery2000_mergermodels}
{Saio} H.,  {Jeffery} C.~S.,  2000, \mn@doi [\mnras] {10.1046/j.1365-8711.2000.03221.x}, \href {https://ui.adsabs.harvard.edu/abs/2000MNRAS.313..671S} {313, 671}

\bibitem[\protect\citeauthoryear{{Saio} \& {Jeffery}}{{Saio} \& {Jeffery}}{2002}]{Saio2002_accretionrcbmodels}
{Saio} H.,  {Jeffery} C.~S.,  2002, \mn@doi [\mnras] {10.1046/j.1365-8711.2002.05384.x}, \href {https://ui.adsabs.harvard.edu/abs/2002MNRAS.333..121S} {333, 121}

\bibitem[\protect\citeauthoryear{{Saumon}, {Chabrier}  \& {van Horn}}{{Saumon} et~al.}{1995}]{Saumon1995}
{Saumon} D.,  {Chabrier} G.,   {van Horn} H.~M.,  1995, \mn@doi [\apjs] {10.1086/192204}, \href {https://ui.adsabs.harvard.edu/abs/1995ApJS...99..713S} {99, 713}

\bibitem[\protect\citeauthoryear{{Schwab}}{{Schwab}}{2018}]{Schwab2018_hehemergers}
{Schwab} J.,  2018, \mn@doi [\mnras] {10.1093/mnras/sty586}, \href {https://ui.adsabs.harvard.edu/abs/2018MNRAS.476.5303S} {476, 5303}

\bibitem[\protect\citeauthoryear{{Schwab}}{{Schwab}}{2019}]{Schwab2019_2DintoMesaSimulations}
{Schwab} J.,  2019, \mn@doi [\apj] {10.3847/1538-4357/ab425d}, \href {https://ui.adsabs.harvard.edu/abs/2019ApJ...885...27S} {885, 27}

\bibitem[\protect\citeauthoryear{Shen}{Shen}{2019}]{Shen2019_MESASS_dwdmerger}
Shen K.,  2019, Double White Dwarf Mergers, \mn@doi{10.5281/zenodo.2603640}, \url {https://doi.org/10.5281/zenodo.2603640}

\bibitem[\protect\citeauthoryear{{Shen}, {Bildsten}, {Kasen}  \& {Quataert}}{{Shen} et~al.}{2012}]{Shen2012_entropyinjection}
{Shen} K.~J.,  {Bildsten} L.,  {Kasen} D.,   {Quataert} E.,  2012, \mn@doi [\apj] {10.1088/0004-637X/748/1/35}, \href {https://ui.adsabs.harvard.edu/abs/2012ApJ...748...35S} {748, 35}

\bibitem[\protect\citeauthoryear{{Timmes} \& {Swesty}}{{Timmes} \& {Swesty}}{2000}]{Timmes2000}
{Timmes} F.~X.,  {Swesty} F.~D.,  2000, \mn@doi [\apjs] {10.1086/313304}, \href {https://ui.adsabs.harvard.edu/abs/2000ApJS..126..501T} {126, 501}

\bibitem[\protect\citeauthoryear{{Tisserand} et~al.,}{{Tisserand} et~al.}{2009}]{Tisserand2009_MagellanicRCBsandDYPers}
{Tisserand} P.,  et~al., 2009, \mn@doi [\aap] {10.1051/0004-6361/200911808}, \href {https://ui.adsabs.harvard.edu/abs/2009A&A...501..985T} {501, 985}

\bibitem[\protect\citeauthoryear{{Tisserand} et~al.,}{{Tisserand} et~al.}{2022}]{Tisserand2022_dLHdC}
{Tisserand} P.,  et~al., 2022, \mn@doi [\aap] {10.1051/0004-6361/202142916}, \href {https://ui.adsabs.harvard.edu/abs/2022A&A...667A..83T} {667, A83}

\bibitem[\protect\citeauthoryear{{Tisserand}, {Crawford}, {Soon}, {Clayton}, {Ruiter}  \& {Seitenzahl}}{{Tisserand} et~al.}{2023}]{Tisserand2023_3Ddistgaia_ARXIVVERSION}
{Tisserand} P.,  {Crawford} C.~L.,  {Soon} J.,  {Clayton} G.~C.,  {Ruiter} A.~J.,   {Seitenzahl} I.~R.,  2023, \mn@doi [arXiv e-prints] {10.48550/arXiv.2309.10148}, \href {https://ui.adsabs.harvard.edu/abs/2023arXiv230910148T} {p. arXiv:2309.10148}

\bibitem[\protect\citeauthoryear{{Townsend} \& {Teitler}}{{Townsend} \& {Teitler}}{2013}]{Townsend2013_gyre}
{Townsend} R.~H.~D.,  {Teitler} S.~A.,  2013, \mn@doi [\mnras] {10.1093/mnras/stt1533}, \href {https://ui.adsabs.harvard.edu/abs/2013MNRAS.435.3406T} {435, 3406}

\bibitem[\protect\citeauthoryear{{Townsend}, {Goldstein}  \& {Zweibel}}{{Townsend} et~al.}{2018}]{Townsend2018_gyre}
{Townsend} R.~H.~D.,  {Goldstein} J.,   {Zweibel} E.~G.,  2018, \mn@doi [\mnras] {10.1093/mnras/stx3142}, \href {https://ui.adsabs.harvard.edu/abs/2018MNRAS.475..879T} {475, 879}

\bibitem[\protect\citeauthoryear{Virtanen et~al.,}{Virtanen et~al.}{2020}]{scipy2020}
Virtanen P.,  et~al., 2020, \mn@doi [Nature Methods] {10.1038/s41592-019-0686-2}, \href {https://rdcu.be/b08Wh} {17, 261}

\bibitem[\protect\citeauthoryear{{Warner}}{{Warner}}{1967}]{1967MNRAS.137..119W}
{Warner} B.,  1967, \mn@doi [\mnras] {10.1093/mnras/137.2.119}, \href {https://ui.adsabs.harvard.edu/abs/1967MNRAS.137..119W} {137, 119}

\bibitem[\protect\citeauthoryear{{Webbink}}{{Webbink}}{1984}]{Webbink1984_dwdmergerscenario}
{Webbink} R.~F.,  1984, \mn@doi [\apj] {10.1086/161701}, \href {https://ui.adsabs.harvard.edu/abs/1984ApJ...277..355W} {277, 355}

\bibitem[\protect\citeauthoryear{{Weiss}}{{Weiss}}{1987}]{Weiss1987_reallyoldRCBevolution}
{Weiss} A.,  1987, \aap, \href {https://ui.adsabs.harvard.edu/abs/1987A&A...185..165W} {185, 165}

\bibitem[\protect\citeauthoryear{{Wong} \& {Bildsten}}{{Wong} \& {Bildsten}}{2024}]{WongBildsten2024_rcbgyre}
{Wong} T. L.~S.,  {Bildsten} L.,  2024, \mn@doi [\apj] {10.3847/1538-4357/ad0cfa}, \href {https://ui.adsabs.harvard.edu/abs/2024ApJ...962...20W} {962, 20}

\bibitem[\protect\citeauthoryear{{Yakovina}, {Pugach}  \& {Pavlenko}}{{Yakovina} et~al.}{2009}]{Yakovina2009_dyper}
{Yakovina} L.~A.,  {Pugach} A.~F.,   {Pavlenko} Y.~V.,  2009, \mn@doi [Astronomy Reports] {10.1134/S1063772909030019}, \href {https://ui.adsabs.harvard.edu/abs/2009ARep...53..187Y} {53, 187}

\bibitem[\protect\citeauthoryear{{Za{\v{c}}s}, {Mondal}, {Chen}, {Pugach}, {Musaev}  \& {Alksnis}}{{Za{\v{c}}s} et~al.}{2007}]{Zacs2007_dyper}
{Za{\v{c}}s} L.,  {Mondal} S.,  {Chen} W.~P.,  {Pugach} A.~F.,  {Musaev} F.~A.,   {Alksnis} O.,  2007, \mn@doi [\aap] {10.1051/0004-6361:20066923}, \href {https://ui.adsabs.harvard.edu/abs/2007A&A...472..247Z} {472, 247}

\bibitem[\protect\citeauthoryear{{Zhang}, {Jeffery}, {Chen}  \& {Han}}{{Zhang} et~al.}{2014}]{Zhang2014_rcbmodels}
{Zhang} X.,  {Jeffery} C.~S.,  {Chen} X.,   {Han} Z.,  2014, \mn@doi [\mnras] {10.1093/mnras/stu1741}, \href {https://ui.adsabs.harvard.edu/abs/2014MNRAS.445..660Z} {445, 660}

\makeatother
\end{thebibliography}


\bsp	
\label{lastpage}
\end{document}